# IceCube: An Instrument for Neutrino Astronomy

Francis Halzen[1] and Spencer R. Klein[2,3]

[1]Department of Physics, University of Wisconsin, 1150 University Avenue, Madison WI 53706
[2]Nuclear Science Division, Lawrence Berkeley National Laboratory, Berkeley, CA, 94720
[3]Department of Physics, University of California, Berkeley, CA, 94720

### Abstract

Neutrino astronomy beyond the Sun was first imagined in the late 1950s; by the 1970s, it was realized that kilometer-scale neutrino detectors were required. The first such instrument, IceCube, is near completion and taking data. The IceCube project transforms a cubic kilometer of deep and ultra-transparent Antarctic ice into a particle detector. A total of 5,160 optical sensors are embedded into a gigaton of Antarctic ice to detect the Cherenkov light emitted by secondary particles produced when neutrinos interact with nuclei in the ice. Each optical sensor is a complete data acquisition system, including a phototube, digitization electronics, control and trigger systems and LEDs for calibration. The light patterns reveal the type (flavor) of neutrino interaction and the energy and direction of the neutrino, making neutrino astronomy possible. The scientific missions of IceCube include such varied tasks as the search for sources of cosmic rays, the observation of Galactic supernova explosions, the search for dark matter, and the study of the neutrinos themselves. These reach energies well beyond those produced with accelerator beams.

The outline of this review is as follows:

- Neutrino Astronomy and Kilometer-Scale Detectors
- High-Energy Neutrino Telescopes: Methodologies of Neutrino Detection
- IceCube Hardware
- High-Energy Neutrino Telescopes: Beyond Astronomy
- Future Projects

## I. Introduction

### The Technology

Soon after the 1956 observation of the neutrino[1], the idea emerged that it represented the ideal astronomical messenger. Neutrinos travel from the edge of the Universe essentially without absorption and with no deflection by magnetic fields. Having essentially no mass and no electric charge, the neutrino is similar to the photon, except for one important attribute: its interactions with matter are extremely feeble. So, high-energy neutrinos may reach us unscathed from cosmic distances, from the inner neighborhood of black holes, and, hopefully, from the nuclear furnaces where cosmic rays are born. Also, in contrast to photons, neutrinos are an unambiguous signature of hadronic interactions.

Their weak interactions make cosmic neutrinos very difficult to detect. Immense particle detectors are required to collect cosmic neutrinos in statistically significant numbers[2]. By the 1970s, it was clear that a cubic-kilometer detector was needed to observe cosmic neutrinos produced by interactions of cosmic rays with background microwave photons[3]. Newer estimates for observing potential cosmic accelerators such as quasars or gamma-ray bursts unfortunately point to the same exigent requirement[4]. Building a neutrino telescope has been a daunting technical challenge.

Given the detector's required size, early efforts concentrated on transforming large volumes of natural water into Cherenkov detectors that catch the light produced when neutrinos interact with nuclei in or near the detector[5]. Building the Deep Underwater Muon and Neutrino Detector (DUMAND) in the sea off the main island of Hawaii unfortunately failed after a two-decade-long effort [6]. However, DUMAND paved the way for later efforts by pioneering many of the detector technologies in use today, and by inspiring the deployment of a smaller instrument in Lake Baikal[7], as well as efforts to commission neutrino telescopes in the Mediterranean[8-10]. The first telescope on the scale envisaged by the DUMAND collaboration was realized instead by transforming a large volume of the extremely transparent, natural deep Antarctic ice into a particle detector, the Antarctic Muon and Neutrino Detector Array (AMANDA). In operation from 2000 to 2009, it represented a proof of concept for the kilometer-scale neutrino observatory, IceCube, which is the main focus of this article[11-12].

Even extremely high-energy neutrinos will routinely stream through a detector without leaving a trace; the few that interact with a nucleus in the ice create muons as well as electromagnetic and hadronic secondary particle showers. The charged secondary particles radiate Cherenkov light that spreads through the transparent ice characterized by an absorption length of 100 m or more, depending on depth. The light pattern reveals the direction of the neutrino, making neutrino astronomy possible. Secondary muons are of special interest, because their mean free path can reach 10 km for the most energetic neutrinos. The effective detector volume thus exceeds the instrumented volume for muon neutrinos. The method is illustrated in Fig.1a.

Photomultipliers transform the Cherenkov light from neutrino interactions into electrical signals using the photoelectric effect. These signals are captured by computer chips that digitize the shape of the current pulses. The information is sent to the computers collecting the data, first by cable to the "counting house" at the surface of the ice sheet and then via magnetic tape. More interesting events are sent by satellite to the IceCube Data Warehouse in Madison, Wisconsin. Essentially, IceCube consists of 5,160 freely running sensors sending time-stamped, digitized waveforms of the light they detect to the surface. The local clocks in the sensors are kept calibrated with nanosecond precision. This information allows the scientists to reconstruct neutrino events and infer their arrival directions and energies.

The complete IceCube detector will observe several hundred neutrinos per day[13-14], with energies above 100 GeV; the DeepCore infill array will identify a smaller sample with energies as low as 10 GeV. These "atmospheric neutrinos" come from the decay of pions and kaons produced by collisions of cosmic-ray particles with nitrogen and oxygen in the atmosphere. Atmospheric neutrinos are a background for cosmic neutrinos, at least for energies below 1,000 TeV, but their flux is calculable and can be used to prove that the detector is performing as expected. At the highest energies, a small charm component is anticipated; its magnitude is uncertain and remains to be measured. As in conventional astronomy, IceCube looks beyond the atmosphere for cosmic signals.

(a)

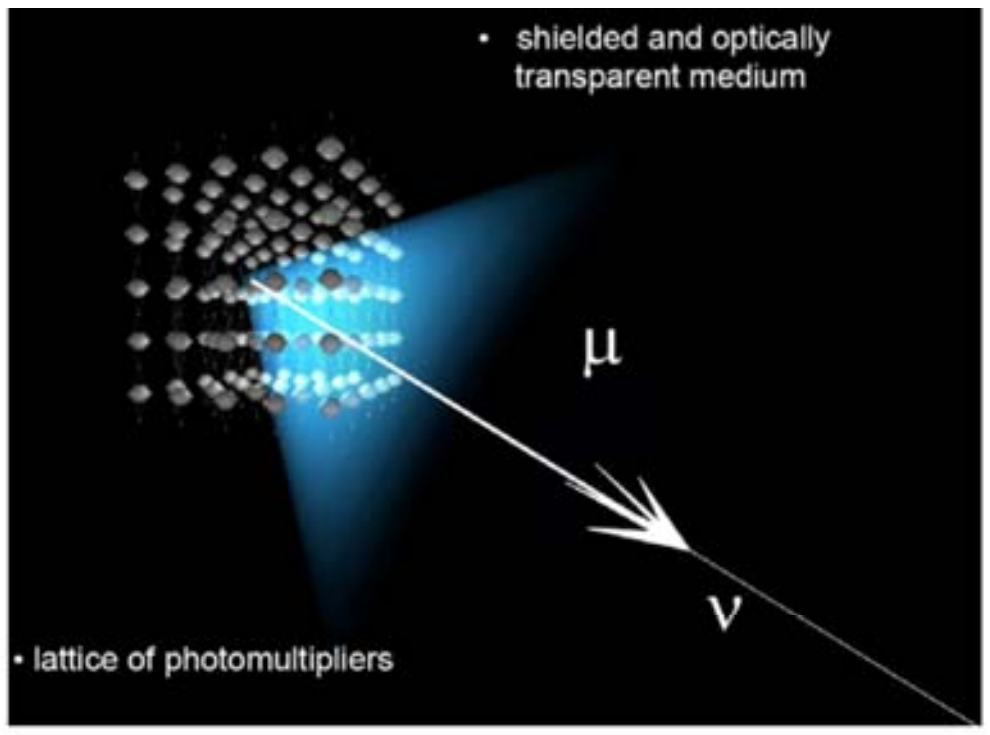

(b)

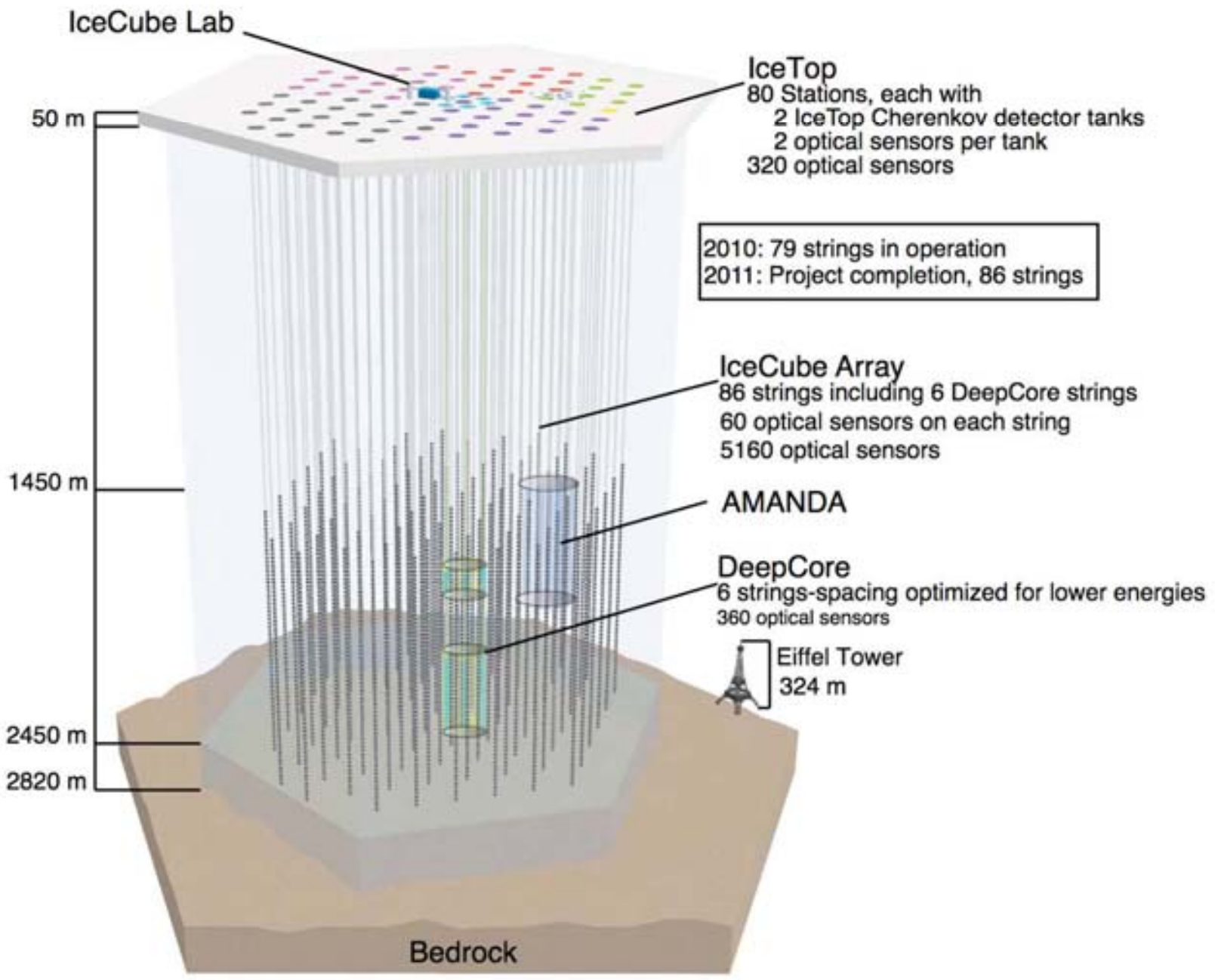

**Fig.1 (Top).** Conceptual design of a large neutrino detector. A neutrino, selected by the fact that it traveled through the Earth, interacts with a nucleus in the ice and produces a muon that is detected by the wake of Cherenkov photons it leaves inside the detector. A high-energy neutrino has a reduced mean free path ($\lambda_\nu$), and the secondary muon an increased range ($\lambda_\mu$), so the probability for observing a muon, $\lambda_\mu/\lambda_\nu$, increases with energy; it is about $10^{-6}$ for a 1-TeV neutrino[15]. (Bottom) Actual design of the IceCube neutrino detector with 5,160 optical sensors viewing a kilometer cubed of natural ice. The signals detected by each sensor are transmitted to the surface over the 86 cables to which the sensors are attached. IceCube encloses it's smaller predecessor, AMANDA.

In parallel, the development of the technology for commissioning a large detector deployed in sea or fresh water (in a lake) continued[16]. Water can have excellent optical quality, with a long scattering length that can lead to very good angular resolution. The decay of radioactive potassium-40 typically contributes a steady 40-kHz background rate in a 10-inch photo multiplier tube (PMT). Bioluminescence also contributes bursts of background light that result in detector deadtime. Currents are also an issue; it is necessary to track the position of the optical sensors. In contrast, Antarctic ice has a shorter scattering length than water, but the attenuation length is longer. With appropriate reconstruction algorithms, it is possible to place the optical sensors farther apart in ice than in water. Furthermore, the only background in the sterile ice is that introduced by the detector itself.

As has already been mentioned, the original effort to build a large detector was by the DUMAND collaboration[6]. They proposed to build a substantial deep-ocean detector at a site about 40 km off the coast of the island of Hawaii, in 4,800 m of water. Buoyant strings of PMTs were to be anchored to the seabed, and connected to the shore by an underwater cable. The challenges were formidable for 1980s technology: high pressures, corrosive salt water and large backgrounds from bioluminescence and radioactive $^{40}K$. DUMAND was cancelled after a pressure vessel leaked during the very first string deployment.

Another effort by a Russian and German collaboration in Lake Baikal, in Siberia, is still taking data[7], taking advantage of the deep, pure water. The detector was built in stages, starting with 36 optical modules; the current ‘main’ detector consists of 192 phototubes on eight strings. A later extension added three ‘sparse’ strings 200 m from the main detector, providing an instrumented mass of 10 megatons for extremely high-energy cascades. The ice that covers Lake Baikal for two months every spring is a convenient platform for detector construction and repair.

After extensive research and development efforts by the ANTARES (Astronomy with a Neutrino Telescope and Abyss Environmental Research), NESTOR (Neutrino Extended Submarine Telescope with Oceanographic Research) and NEMO (Neutrino Mediterranean Observatory) collaborations in the Mediterranean, there is optimism that the technological challenges to building neutrino telescopes in deep seawater have now been met[16]. Construction of the ANTARES detector, located at a depth of 2,400 m close to the shore near Toulon, France, has been completed[9]. The detector consists of 12 strings, each equipped with 75 optical sensors mounted in 25 triplets. ANTARES’ performance has been verified by the first observation of atmospheric neutrinos. Like AMANDA, it is a proof of concept for KM3NeT, a kilometer-scale detector in the Mediterranean Sea, complementary to IceCube at the South Pole.

**The Science**

Neutrino astronomy predates kilometer-scale detectors[17]. The first searches for extra-terrestrial neutrinos were in the 1960s, in two deep mines: India’s Kolar Gold Field and South Africa’s East Rand mine. Because of the large background radiation at ground level, from cosmic rays interacting in the atmosphere, neutrino detectors must be underground. Both experiments used scintillation detectors a few meters on each side to detect a handful of upward-going muons from atmospheric neutrinos. By 1967, Ray Davis’ geochemical experiment was detecting a few argon atoms a day, produced when solar neutrinos interacted in an underground tank filled with perchloroethylene[18].

By the late 1980s, scintillation detectors had evolved into the 78-meters long by 12-meters wide by 9-meters-high MACRO detector in the Gran Sasso underground laboratory in Italy. MACRO consisted of passive absorber interspersed with streamer tubes, and surrounded by 12 meter long tanks containing liquid scintillator[18-19]. MACRO observed over 1,000 neutrinos over the course of 6 years. In a similar period, the Frejus experiment measured the atmospheric $\nu_\mu$ spectrum and set a limit on TeV extra-terrestrial neutrinos[21]. However, further growth required a new technique, first suggested by Markov in 1960: detecting charged particles by the Cherenkov radiation emitted in water or ice[5].

Cherenkov light is radiated by charged particles moving faster than the speed of light in the medium; in ice, this is 75% of the speed of light in a vacuum. The emission is akin to a sonic boom. Photomultiplier tubes (PMTs) detect this blue and near-UV light. With a sufficient density of PMTs, neutrinos with energies of only a few MeV may be reconstructed. The water Cherenkov technique was pioneered in kiloton-sized detectors, optimized for relatively low-energy (GeV) neutrinos. The two most successful first-generation detectors were the Irvine-Michigan-Brookhaven[22] and Kamiokande[23] detectors. Both consisted of tanks containing thousands of tons of purified water, monitored with 1,000s of PMTs on the top and sides of the tank. Although optimized for GeV energies, these detectors were also sensitive to lower energy neutrinos; IMB[22] and Kamiokande[23] launched neutrino astronomy by detecting some 20 low-energy (10-50 MeV) neutrino events from supernova 1987A.

Their success, as well as the accumulating evidence for the “solar neutrino puzzle”, stimulated the development of two second-generation detectors. Super-Kamiokande is a 50,000-ton, scaled-up version of Kamiokande[24], and the Sudbury Neutrino Observatory (SNO) is a 1,000-ton, heavy-water ($D_2O$)-based detector[25]. Together, the two experiments clearly showed that neutrinos have mass by observing flavor oscillations (between $\nu_\mu$, $\nu_e$ and $\nu_\tau$) in the solar and atmospheric-neutrino beams, thus providing the first evidence for physics beyond the Standard Model. These experiments showed that at GeV energies, atmospheric neutrinos were a major background to searches for non-thermal astronomical sources where particles, e.g. the observed cosmic rays, are accelerated. The spectrum of cosmic neutrinos from these sources extends to energies beyond those characteristic of atmospheric neutrinos. Future experiments would require kilometer-scale volumes, and would target higher energies where the background is lower. Although Super-Kamiokande continues to collect data, there is considerable interest in building much-larger megaton detectors to pursue these physics studies with higher sensitivity.

In summary, the field has already achieved spectacular success: neutrino detectors have “seen” the Sun and detected a supernova in the Large Magellanic Cloud in 1987. Both observations were of tremendous importance; the former showed that neutrinos have a tiny mass, opening the first crack in the Standard Model of Particle Physics, and the latter confirmed the theory of stellar evolution as well as the basic nuclear physics of the death of stars. Figure 2 illustrates the cosmic-neutrino energy spectrum covering an enormous range, from microwave energies ($10^{-12}$ eV) to $10^{20}$ eV[26]. The figure is a mixture of observations and theoretical predictions. At low energy, the neutrino sky is dominated by neutrinos produced in the Big Bang. At MeV energy, neutrinos are produced by supernova explosions; the flux from the 1987 event is shown. The figure displays the measured atmospheric-neutrino flux up to energies of 100 TeV by the AMANDA experiment[27]. Atmospheric neutrinos are a key to our story, because they are the dominant background for extra-terrestrial searches. The flux of atmospheric neutrinos falls dramatically with increasing energy; events above 100 TeV are rare, leaving a clear field of view for extra-terrestrial sources.

The highest-energy neutrinos in Fig.2 are the decay products of pions produced by the interactions of cosmic rays with microwave photons[28]. Above a threshold of $\sim 4\times10^{19}$ eV, cosmic rays interact with the microwave background, introducing an absorption feature in the cosmic-ray flux, the Greissen-Zatsepin-Kuzmin (GZK) cutoff. As a consequence, the mean free path of extragalactic cosmic rays propagating in the microwave background is limited to roughly 75 megaparsecs (240 million light years) and, therefore, the secondary neutrinos are the only probe of the still-enigmatic sources at longer distances. What they will reveal is a matter of speculation. The calculation of the neutrino flux associated with the observed flux of extragalactic cosmic rays is straightforward, and yields one event per year in a kilometer-scale detector. It is however subject to ambiguities, most notably from the still-unknown composition of the highest-energy cosmic rays, and due to the cosmological evolution of the sources[29]. The flux, labeled GZK in Fig.2, shares the high-energy neutrino sky with neutrinos from gamma-ray bursts and active galactic nuclei[4].

In this review, we will first illustrate the origin of the concept to build a kilometer-scale neutrino detector. It has taken half a century from the concept to the commissioning of IceCube. It took this long to develop the methodologies and technologies to build a neutrino telescope; we will describe them next. We complete the article by discussing other science covered by this novel instrument.

## II. Why Kilometer-Scale Detectors: Neutrino Sources and Cosmic Rays

### Cosmic-Ray Accelerators and Cosmic-Beam Dumps

Despite a discovery potential touching a wide range of scientific issues, construction of IceCube and a future KM3NeT[10] has been largely motivated by the possibility of opening a new window on the Universe, using neutrinos as cosmic messengers. Specifically, we will revisit IceCube's prospects to detect cosmic neutrinos associated with cosmic rays, and thus finally reveal their sources.

Cosmic accelerators produce particles with energies in excess of $10^8$ TeV; we still do not know where or how[30]. The observed flux of cosmic rays is shown in Fig.3[26]. The energy spectrum follows a sequence of three power laws. The first two are separated by a feature dubbed the “knee" at an energy of approximately 3,000 TeV. There is evidence that cosmic rays up to this energy are Galactic in origin. Any association with our Galaxy disappears in the vicinity of a second feature in the spectrum referred to as the “ankle"; see Fig.3. Above the ankle, the gyroradius of a proton in the Galactic magnetic field exceeds the size of the Galaxy, and points to the onset of an extragalactic component in the spectrum that extends to energies beyond $10^8$

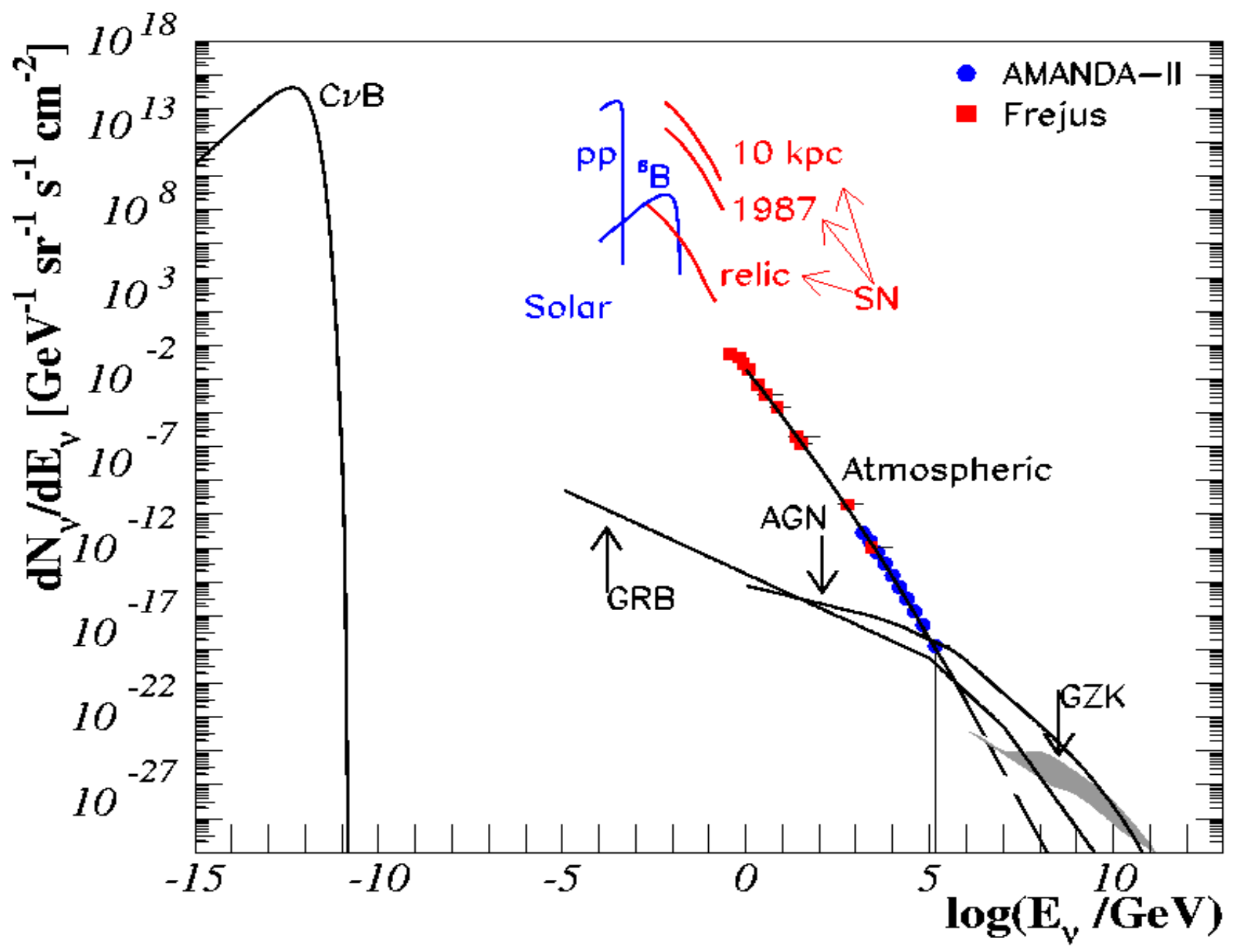

**Fig.2** The cosmic-neutrino spectrum. Sources are the big bang (CνB), the Sun, supernovae (SN), atmospheric neutrinos, active galactic nuclei (AGN) galaxies, and GZK neutrinos. The data points are from detectors at the Frejus underground laboratory[21] (red) and from AMANDA[27] (blue).

TeV. Direct support for this assumption comes from two experiments that have observed the telltale structure in the cosmic-ray spectrum resulting from the absorption of the particle flux by the microwave background, the so-called GZK cutoff. The origin of the flux in the intermediate region remains a mystery, although it is routinely assumed that it results from some high-energy extension of the reach of Galactic accelerators.

Acceleration of protons (or nuclei) to TeV energy and above likely requires massive bulk flows of relativistic charged particles. These are likely to originate from exceptional gravitational forces in the vicinity of black holes or neutron stars. Gravity powers large currents of charged particles that produce high magnetic fields. These fields create the opportunity for particle acceleration by shocks, similar to what happens with solar flares. It is a fact that electrons are accelerated to TeV energy and above near black holes; astronomers detect them indirectly by their synchrotron radiation. Some must accelerate protons, because we observe them as cosmic rays.

How many gamma rays and neutrinos are produced in association with the cosmic-ray beam? Generically, a cosmic-ray source should also be a “beam dump”. Cosmic rays accelerated in regions of high magnetic fields near black holes inevitably interact with radiation surrounding them: for instance, UV photons in active galaxies or MeV photons in gamma-ray-burst fireballs. Neutral and charged pion secondaries are produced by the processes

$$p+\gamma \rightarrow \pi^{o} + p \ \text{ and } \ p+\gamma \rightarrow \pi^{+} + n. \qquad (1)$$

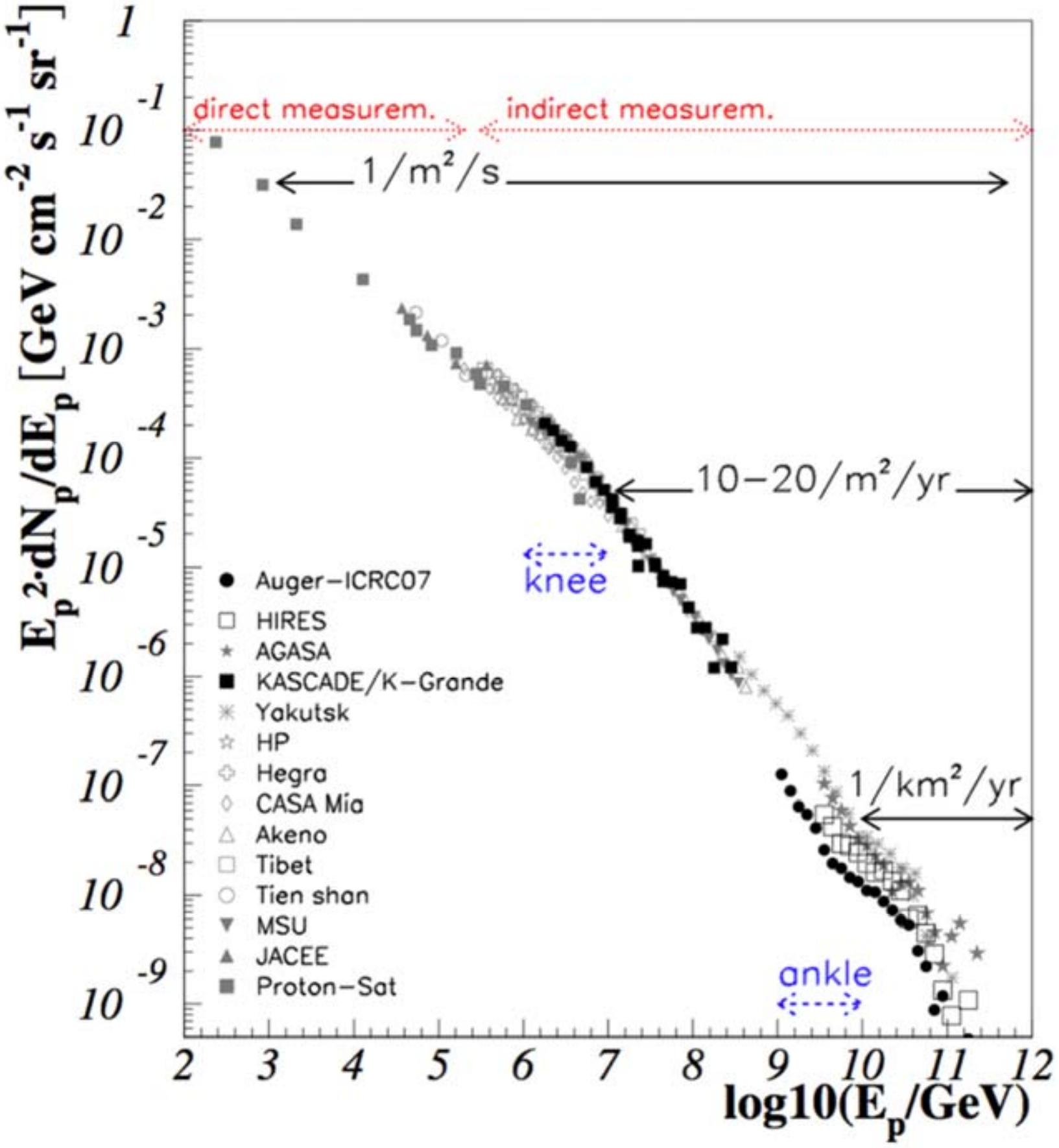

**Fig.3** At the energies of interest here, the cosmic-ray spectrum follows a sequence of 3 power laws. The first 2 are separated by the "knee", the 2nd and 3rd by the "ankle". Cosmic rays beyond the ankle are a new population of particles produced in extragalactic sources[26].

Although secondary protons may remain trapped in the high magnetic fields, neutrons and the pion decay products escape. The energy escaping the source is distributed among cosmic rays, gamma rays and neutrinos produced by the decay of neutrons, neutral pions and charged pions, respectively. Kilometer-scale neutrino detectors have the sensitivity to reveal generic cosmic-ray sources with an energy density in neutrinos comparable to their energy density in cosmic rays and pionic TeV photons[31].

In the case of Galactic supernova shocks, cosmic rays interact with gas in the Galactic disk, e.g. with dense molecular clouds, producing equal numbers of pions of all three charges in hadronic collisions $p + p \rightarrow n\left[\pi^o + \pi^+ + \pi^-\right] + X$. Here n is the multiplicity of secondary pions.

This mechanism predicts a relation between cosmic-ray ($N_p$), gamma-ray ($N_\gamma$) and neutrino ($N_\nu$) fluxes[31]

$$\frac{dN_\nu}{dE} = \frac{1}{2}\left[\frac{1}{8}\right]\frac{dN_\gamma}{dE}, \tag{2}$$

$$\frac{dN_\nu}{dE_\nu} \cong n_{\rm int} x_\nu \frac{dN_p}{dE_p}\left(\frac{E_p}{x_\nu}\right). \quad (3)$$

The first relation reflects the fact that pions decay into gamma rays and neutrinos that carry 1/2 and 1/4 of the energy of the parent. This assumes that the four leptons in the decay $\pi^+ \rightarrow \nu_\mu + (e + \bar{\nu}_e + \nu_\mu)$ equally share the charged pion's energy. $N_\nu (= N_{\nu_\mu} = N_{\nu_e} = N_{\nu_\tau})$ is the sum of the neutrino and antineutrino fluxes which are not distinguished by the experiments. Oscillations over cosmic baselines yield approximately equal fluxes for the three flavors. The two factors apply to the hadronic and photoproduction of pions in the source, respectively. Although this relation only depends on straightforward particle physics, the second relation of the neutrino to the actual cosmic-ray flux depends on $n_{int}$, the number of interactions determined by the optical depth of the source for p$\gamma$ interactions. The factor

$$x_\nu = \frac{E_\nu}{E_p} = \frac{1}{4}\langle x_{p\rightarrow\pi}\rangle \cong \frac{1}{20}, \quad (4)$$

is the relative energy of the neutrino to the pion. The pion carries, on average, a fraction $x_{p\rightarrow\pi} \sim 0.2$ of the parent proton energy and shares it roughly equally between the 4 leptons.

These relations form the basis for testing the assumption that cosmic rays are accelerated in a cosmic source. For a more detailed discussion of these relations, we refer the reader to reference[31].

This discussion does not apply to sources that primarily accelerate electrons, which do not produce neutrinos. Some cosmic electron accelerators have been identified via their emission of polarized synchrotron radiation. However, unambiguously identifying a source that does not emit synchrotron radiation is challenging, and unambiguous observation of a cosmic-ray accelerator may require the observation of neutrinos.

### Galactic Sources

Supernova remnants were proposed as the source of Galactic cosmic rays as early as 1934 by Baade and Zwicky[32]; their proposal is still a matter of debate. Galactic cosmic rays reach energies of at least several thousand TeV, the “knee" in the spectrum. Their interactions with Galactic hydrogen in the vicinity of the accelerator should generate gamma rays from decay of secondary pions that reach energies of hundreds of TeV. Such sources should be identifiable by a relatively flat energy spectrum that extends to high energy without attenuation; they have been dubbed PeVatrons. Straightforward energetics arguments show that present air Cherenkov telescopes should have the sensitivity necessary to detect TeV photons from PeVatrons.

They may have been revealed by an all-sky survey in ~10 TeV gamma rays with the Milagro detector[33]. Sources are identified in nearby star-forming regions in Cygnus and in the vicinity of Galactic latitude $l$ = 40 degrees; some are not readily associated with known supernova remnants or with non-thermal sources observed at other wavelengths. In fact, some Milagro sources may actually be molecular clouds illuminated by the cosmic-ray beam accelerated in young remnants located within ~100 pc. One expects indeed that the highest-energy cosmic rays are accelerated over a short time period, of order one to ten thousand years when the shock velocity is high. The high-energy particles can produce photons and neutrinos over much longer periods when they diffuse through the interstellar medium to interact with nearby molecular clouds[34]. Star-forming regions provide all ingredients for the efficient production of neutrinos: supernovae to accelerate cosmic rays and a high density ambient medium, including molecular clouds, as an efficient target for producing pions.

Despite the rapid development of more sensitive instruments, it has been impossible to conclusively pinpoint supernova remnants as sources of cosmic rays by identifying gamma rays of pion origin. Eliminating the possibility of a purely electromagnetic origin of TeV gamma rays is challenging. Detecting the accompanying neutrinos would provide incontrovertible evidence for cosmic-ray acceleration in the sources. Particle physics dictates the relation between gamma rays and neutrinos, and basically predicts the production of a $\nu_\mu + \bar{\nu}_\mu$ pair for every two gamma rays seen by Milagro; see Eq. 2. This follows from the assumptions that gamma rays and neutrinos originate indeed from pions produced in equal numbers for each of the three charges in the collisions of cosmic rays with the interstellar matter. For average values of the parameters describing the flux, the completed IceCube detector could confirm the sources in the Milagro sky map as sites of cosmic-ray acceleration at the 3σ level in less than 1 year and at the 5σ level in 3 years[35-36]; see also Fig.4.

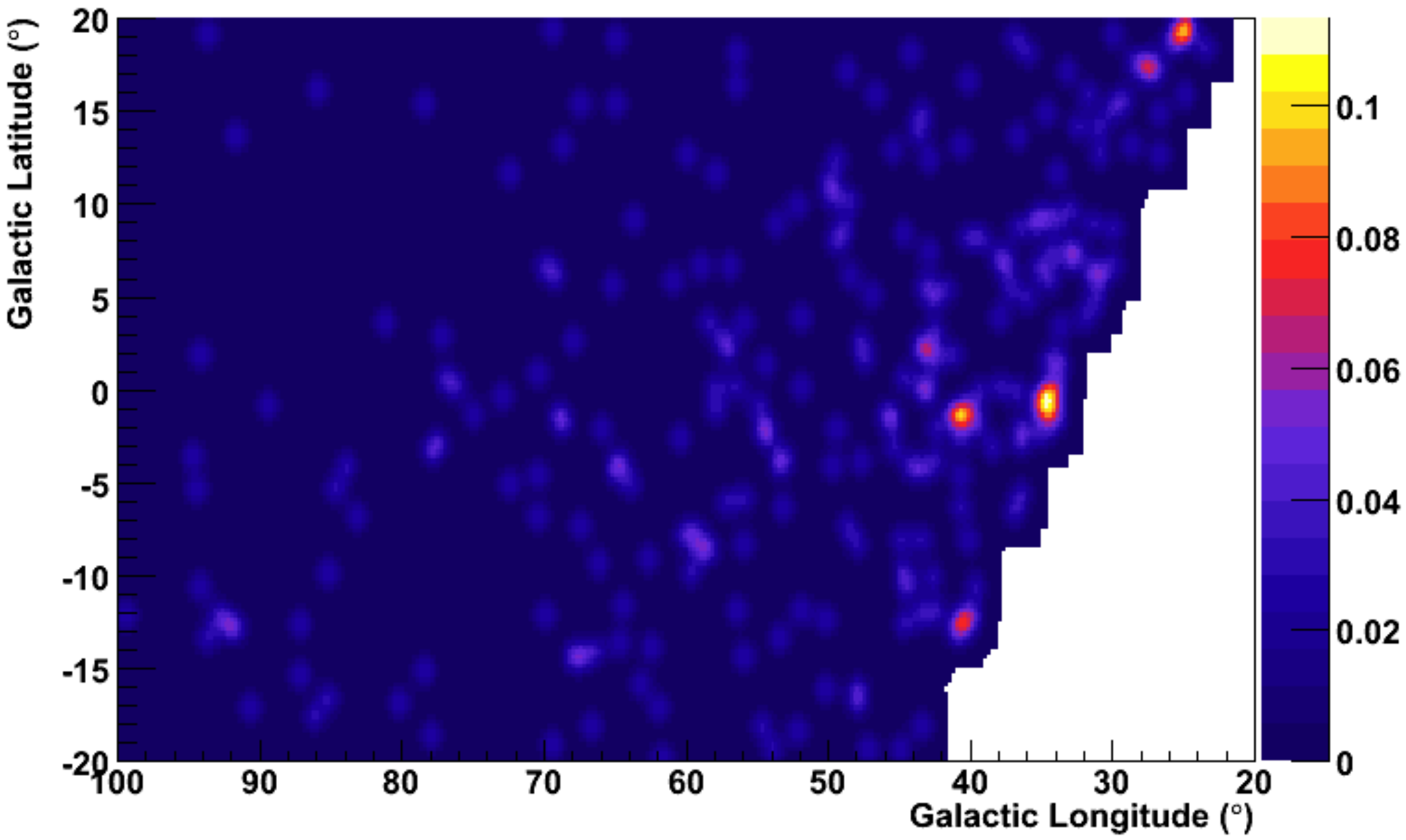

**Fig.4** Simulated sky map from IceCube in Galactic coordinates after 5 years of operation of the completed detector. Two Milagro sources are visible "by eye" with 4 events for MGRO J1852+01 and 3 for MGRO J1908+06 with energy in excess of 40 TeV. These, along with background events, have been randomly distributed according to the resolution of the detector and the size of the sources.

## Sources of Extragalactic Cosmic Rays

Although there is no direct evidence that supernovae accelerate cosmic rays, the idea is generally accepted because of energetics: three supernovae per century converting a reasonable fraction of a solar mass into particle acceleration can accommodate the steady flux of cosmic rays in the Galaxy. Energetics also drives speculation on the origin of extragalactic cosmic rays. The energy density of these rays in the Universe is $\rho_E \simeq 3 \times 10^{-19}$ erg cm$^{-3}$. It can be accommodated with the reasonable assumption that shocks in the expanding gamma-ray-burst (GRB) fireball convert similar energy into the acceleration of protons that are observed as cosmic rays. It so happens that $2 \times 10^{51}$ erg per GRB will yield the observed energy density in cosmic rays after $10^{10}$ years, given that their rate is 300 per (Gpc)$^3$ per year. Therefore, 300 GRBs per year over Hubble time produce the observed cosmic-ray energy density in the Universe, just as three supernovae per century accommodate the steady flux of cosmic rays in the Galaxy[31-32].

Cosmic rays and synchrotron photons coexist in the expanding GRB fireball prior to it reaching transparency and producing the observed GRB display. Their interactions produce charged pions and neutrinos with a flux that can be estimated from the observed extragalactic cosmic-ray flux; see Eq. 3. Fireball phenomenology predicts that, on average, $n_{int} \approx 1$.

Problem solved? Not really: the energy density of extragalactic cosmic rays can also be accommodated by active Galactic nuclei, provided each converts $2 \times 10^{44}$ ergs$^{-1}$ into particle acceleration. As with GRBs, this is an amount that matches their output in electromagnetic radiation[39].

Waxman and Bahcall[40] have argued that it is implausible that the neutrino flux should exceed the cosmic-ray flux

$$E_\nu^2 \frac{dN}{dE_\nu} = 5 \times 10^{-11} \text{ TeV cm}^{-2} \text{ s}^{-1} \text{ sr}^{-1}. \qquad (5)$$

For the specific example of GRB, we have to scale it downward by a factor $x_\nu \approx 1/20$; see Eq. 3. After 7 years of operation, AMANDA's sensitivity is approaching the interesting range, but it takes IceCube to explore it.

If GRB are the sources[41], and the flux is near this limit, then IceCube's mission is relatively straightforward, because we expect to observe of order 10 neutrinos per kilometer square per year in coincidence with GRBs observed by the Swift and Fermi satellites, which translates to a $5\sigma$ observation[42]. Similar statistical power can be obtained by detecting showers produced by $\nu_e$ and $\nu_\tau$.

In summary, while the road to identification of sources of the Galactic cosmic ray has been mapped, the origin of the extragalactic component remains unresolved. Hopefully, neutrinos will reveal the sources.

## III. Neutrino Telescopes: The Concept

Because of the small neutrino cross-sections, a very large detector is required to observe astrophysical neutrinos. At the same time, flavor identification is also very desirable, since the background from atmospheric neutrinos is much lower for $\nu_e$ and $\nu_\tau$ than for $\nu_\mu$. Of course, angular resolution is also very important for detecting point sources, and energy resolution is important in determining neutrino energy spectra, which is important for identifying a diffuse flux of extra-terrestrial neutrinos.

IceCube detects neutrinos by observing the Cherenkov radiation from the charged particles produced by neutrino interactions. Charge-current interactions produce a lepton, which carries an average of 50% (for $E_\nu \sim 10$ GeV) to 80% (at high energies) of the neutrino energy; the remainder of the energy is transferred to the nuclear target. The latter is released in the form of a hadronic shower; both the produced lepton and the hadronic shower produce Cherenkov radiation. In neutral-current interactions, the neutrino transfers a fraction of its energy to a nuclear target, producing just a hadronic shower.

IceCube can differentiate neutrino interactions on the basis of their topology, as is shown in Fig. 5. At low energies, there are two basic topologies: tracks from $\nu_\mu$, and cascades from $\nu_e$, $\nu_\tau$ and all-flavor neutral-current interactions. Charge current cascades include contributions from the shower from the electron (or tau decay products) plus the hadronic shower from the struck nucleus; the contributions are inseparable.

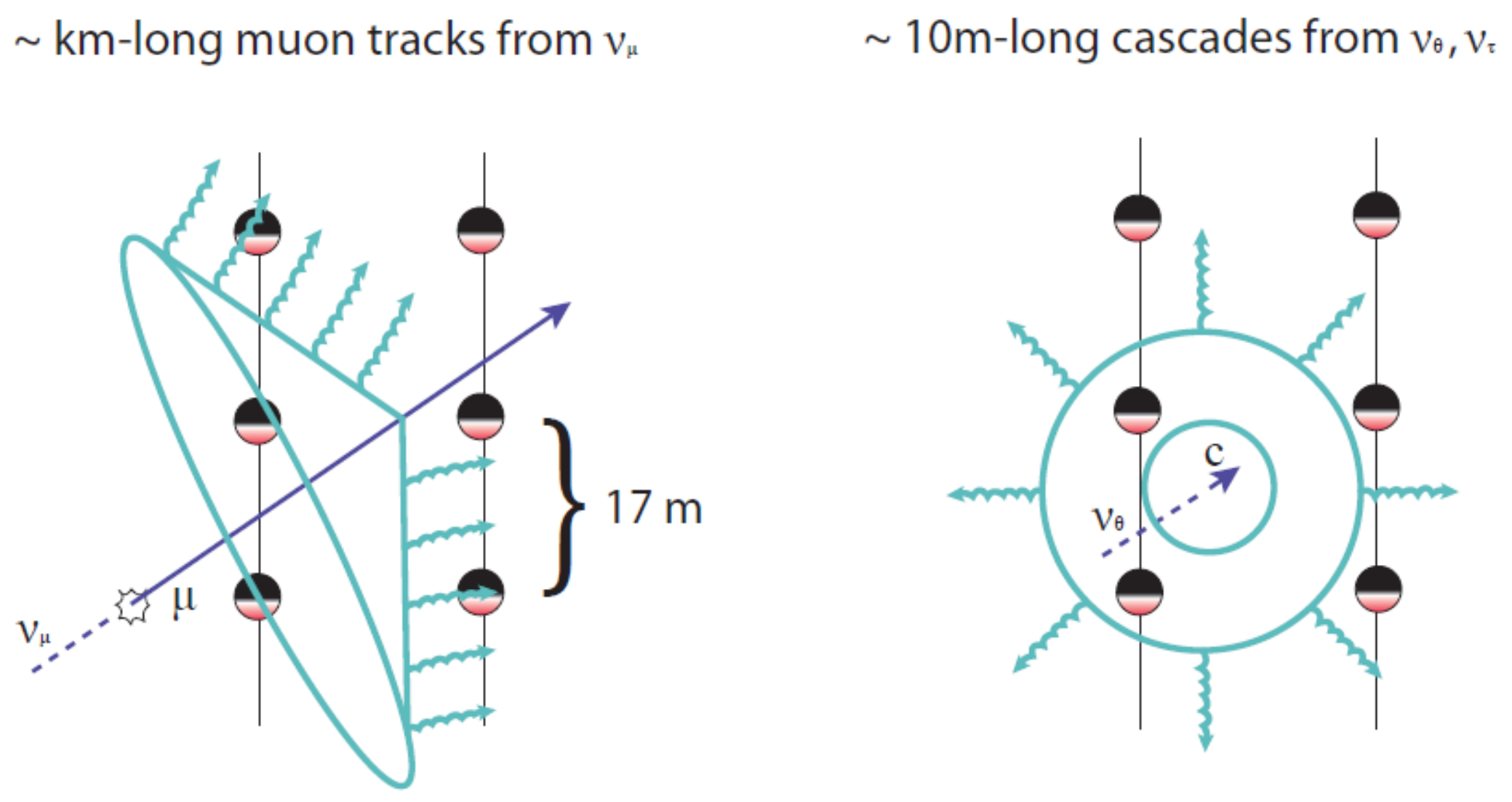

**Fig.5**. Contrasting Cherenkov light patterns produced by muons (left) and by showers initiated by electron and tau neutrinos (right) and by neutral current interactions. The patterns are often referred to as tracks and cascades (or showers). Cascades are produced by a (approximately) point source of light with respect to the dimensions of the detector. At PeV energies, $\tau$ leptons travel hundreds of meters before decaying, producing a third topology, with two cascades – one when the $\nu_\tau$ interacts, and the second when the $\tau$ decays[43]. This is the 'double bang' signature; a simulated event is shown in Fig. 22.

At PeV energies, muon tracks can be up to 10 kilometers long, while, on the scale of IceCube, cascades are nearly point sources. At higher (PeV) energies, an additional topology arises. This is the 'double bang' whereby a $\nu_\tau$ interacts, and the energy transferred to the target nucleus produces one cascade. The $\tau$ travels some hundreds of meters and decays, producing a second cascade.

The different topologies each have advantages and disadvantages. The long lever arm from tracks from $\nu_\mu$ decay allows the muon direction (and, from that, the neutrino direction) to be determined accurately; as will be seen, IceCube's angular resolution is better than 1° for long tracks. One can produce sky maps and search them for hot spots. This is obviously key in finding neutrino sources. The disadvantages are that there is a large background of atmospheric $\nu_\mu$, and that, because the events are not contained, it is difficult to determine the neutrino energy.

However, $\nu_e$ and $\nu_\tau$ interactions also have some significant advantages. They are detected in both the Northern and Southern Hemispheres. (This is also true for $\nu_\mu$ with energy above 1 PeV, where the background from the steeply falling atmospheric spectrum becomes negligible.) IceCube's sensitivity to the Galactic center is similar to that of ANTARES, although not to that of a kilometer-scale detector in the Northern Hemisphere.
The background of atmospheric $\nu_e$ is significantly lower and there are almost no atmospheric $\nu_\tau$. At higher energies muons from $\pi$ decay, the source of atmospheric $\nu_e$, no longer decay and relatively rare K-decays become the dominant source of background $\nu_e$.

The neutrino energy may be more precisely determined. Since the events are contained, the energy measurement is largely calorimetric, because the light output scales nearly linearly with the cascade energy. Their energy measurement is superior. One can use energy spectra to differentiate between atmospheric and extra-terrestrial neutrinos; the latter have a harder spectrum.

Tau neutrinos, $\nu_{\tau,}$ are not absorbed by the Earth[44]. Instead, $\nu_\tau$ interacting in the Earth produce secondary $\nu_\tau$ of lower energy, either directly in a neutral current interaction or via the decay of a tau lepton produced in a charged-current interaction. High-energy $\nu_\tau$ will thus cascade down to hundreds of TeV energy where the Earth is transparent. In other words, they are detected with a reduced energy, but are not absorbed.

Although cascades are nearly point-like, they are not isotropic; light is preferentially emitted at the Cherenkov angle, about 41 degrees in ice. Although this light is heavily scattered before reaching the optical sensors, at energies above 100 TeV enough directional information may remain to determine the neutrino direction to about 30°; see Ref. 45. The light produced by cascades spreads over a large volume; in IceCube, a 10-TeV cascade is visible within a radius of about 130 m rising to 460 m at 10 EeV, i.e. the shower radius grows by just over 50 m per decade in energy.

At energies above about 1 PeV in ice, electrons and photons can interact with multiple atoms, and the Landau-Pomeranchuk-Migdal (LPM) effect reduces the cross-sections for bremsstrahlung and pair production. At energies above about 100 PeV, electromagnetic showers begin to elongate, reaching a length of about 80 meters at 100 EeV[46]. At these energies, the shower direction might be better determined. At energies above about 100 PeV, photonuclear interactions must be considered, and even electromagnetic showers will have a hadronic component, including some muon production.

The detection of neutrinos of all flavors is important in separating diffuse extra-terrestrial neutrinos from atmospheric neutrinos. Generic cosmic accelerators produce neutrinos from the decay of pions with admixture $\nu_e:\nu_\mu:\nu_\tau$ = 1:2:0. Over cosmic baselines, neutrino oscillations alter the ratio to 1:1:1, because approximately one-half of the $\nu_\mu$ convert to $\nu_\tau$. The same production ratio is expected for lower-energy (below 10 GeV) atmospheric neutrinos, where the muons can decay before interacting. However, at higher energies, the muons interact, and atmospheric neutrinos are largely $\nu_\mu$. The flavor ratio depends on the distance the neutrinos have travelled; extra-terrestrial neutrinos should have comparable fluxes of $\nu_e$, $\nu_\mu$ and $\nu_\tau$.

For in-depth discussion of neutrino detection, energy measurement and flavor separation, and for detailed references; see the IceCube Preliminary Design Document[11] and Ref. 15.

## Detection Probabilities

To a first approximation, neutrinos are detected when they interact inside the instrumented volume. The path length L(θ) traversed within the detector volume by a neutrino with zenith angle θ is determined by the detector's geometry. To a first approximation, neutrinos are detected if they interact within the detector volume, i.e. within the instrumented distance L(θ). That probability is

$$P(E_\nu) = 1 - \exp\left(-\frac{L}{\lambda_\upsilon(E_\nu)}\right) \cong \frac{L}{\lambda_\upsilon(E_\nu)} , \qquad (6)$$

where

$$\lambda_\nu(E_\nu) = \left[\rho_{ice} N_A \sigma_{\nu N}(E_\nu)\right]^{-1} \qquad (7)$$

is the mean-free path in ice for a neutrino of energy $E_\nu$. Here $\rho_{ice} = 0.9$ g cm$^{-3}$ is the density of the ice, $N_A = 6.022 \times 10^{23}$ is Avogadro's number and $\sigma_{\nu N}(E_\nu)$ is the neutrino-nucleon cross section. A neutrino flux, $dN/dE_\nu$ (neutrinos per GeV per cm$^2$ per s), crossing a detector with energy threshold $E_\nu^{th}$ and cross-sectional area $A(E_\nu, \theta)$ facing the incident beam will produce

$$N_{ev} = T \int_{E_\nu^{th}} A(E_\nu) P(E_\nu) \frac{dN}{dE_\nu} dE_\nu \qquad (8)$$

events after a time *T*. One must additionally account for the fact that neutrinos may not reach the detector, because they are absorbed in the Earth when they travel along a chord of length X (θ) at zenith angle θ. This absorption factor depends on neutrino flavor and must also be included in the probability $P(E_\nu)$ that the neutrino is detected . Event-rate calculations are discussed in more detail in the appendix of Ref. 36.

So far the formalism applies to contained events, i.e. we assumed that the neutrino interacted within the instrumented distance L(θ). Furthermore, the "effective" detector area $A(E_\nu, \theta)$ is clearly also a function of zenith angle θ. It isn't strictly equal to the geometric cross-section of the instrumented volume facing the incoming neutrino, because even neutrinos interacting outside the instrumented volume may produce enough light inside the detector to be detected. In practice, $A(E_\nu, \theta)$ is determined as a function of the incident neutrino direction and zenith angle by a full-detector simulation, including the trigger. It is of order 1 $km^2$ for IceCube. Often the neutrino effective area is introduced as $A_\nu = AP$ . Note that the quantity P is calculated rather than measured and is different for muon and tau flavors; we generalize it next.

For muon neutrinos, any neutrino producing a secondary muon that reaches the detector (and has sufficient energy to trigger it) will be detected; see Fig.1a. Because the muon travels kilometers at TeV energy and tens of kilometers at PeV energy, neutrinos can be detected outside the instrumented volume; the probability is obtained by substitution in Eq. 6:

$$L \to \lambda_\mu, \qquad (9)$$

therefore,

$$P = \frac{\lambda_\mu}{\lambda_\nu}. \qquad (10)$$

Here, $\lambda_\mu$ is the range of the muon determined by its energy loss.

A tau neutrino can be observed provided the tau lepton it produces reaches the instrumented volume within its lifetime. In Eq. 6, L is replaced by

$$L \to \gamma c\tau = (E/m)c\tau, \qquad (11)$$

where m, τ and E are the mass, lifetime and energy of the tau, respectively. The tau decay length $\lambda_\tau = \gamma c\tau \approx 50m \times (E/10^6 GeV)$ grows linearly with energy and exceeds the range of the muon near $10^{18}$ eV. At even higher energies, the tau eventually ranges out by catastrophic interactions, just like the muon, despite reduction of the cross-sections by a factor $(m_\mu/m_\tau)^2$. The taus trigger the detector but the tracks and (or) showers they produce are mostly indistinguishable from those initiated by muon and electron neutrinos; see also Fig. 5.

To be clearly recognizable as $\nu_\tau$, both the initial neutrino interaction and the subsequent tau decay must be contained within the detector; for a cubic-km detector, this happens for neutrinos with energies from a few PeV to a few 10's of PeV. It might be possible to identify $\nu_\tau$ that only interact in the detector, or τ that decay in the detector[47], but this has not yet been proven.

## Muon Energy measurement

Muons from $\nu_\mu$ have ranges from kilometers at TeV energy to tens of kilometers at EeV energy, generating showers along their track by bremsstrahlung, pair production and photonuclear interactions. These are the sources of additional Cherenkov radiation. Because the energy of the muon degrades along its track, the energy of secondary showers also gradually diminishes, and the distance from the track over which the associated Cherenkov light can trigger a PMT is gradually reduced. The geometry of the light pool surrounding the muon track is therefore a kilometer-long cone with a gradually decreasing radius. At lower energies, of hundreds of GeV and less, the muon becomes minimum-ionizing.

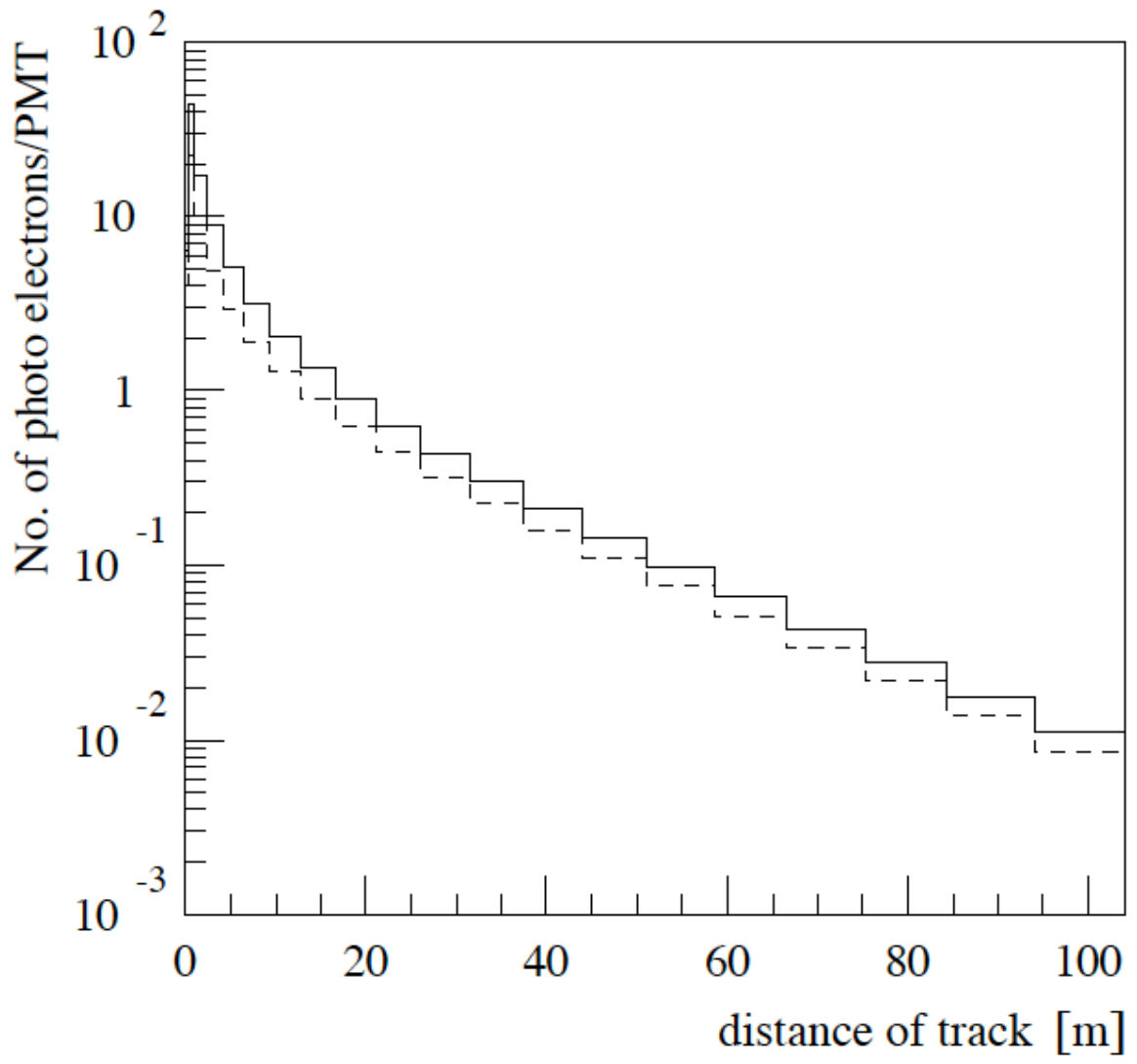

**Fig.6.** The average number of photoelectrons observed by an 8" AMANDA PMT is shown as a function of the minimum distance to a minimum ionizing muon track. The result for the average PMT direction (dashed line) and the direction towards the Cherenkov cone (solid line) are shown. On the average, a minimum ionizing particle is visible up to 20 meters from a PMT.

In its first kilometer, a high-energy muon typically loses energy in a couple of showers of one-tenth of its initial energy. So the initial radius of the cone is the radius of a shower with 10% of the muon energy, e.g. 130 m for a 100-TeV muon. Near the end of its range, the muon becomes minimum-ionizing, visible up to about 15 meters from the PMT. Figure 6 shows the detection distance as a function of photoelectron threshold.

Because of the stochastic nature of the muon's energy loss, the relationship between observed (via Cherenkov light) energy loss and muon energy varies from muon to muon. The muon energy in the detector can be determined to roughly a factor of two. Beyond that, one does not know how far the muon traveled (and how much energy it lost) before entering the detector; an unfolding process is required to determine the neutrino energy based on observed muon energies.

## IV. From AMANDA to IceCube: Natural Antarctic Ice as a Cherenkov Detector

Neutrino detection in ice was pioneered by the AMANDA collaboration in the late 1990s[48]. It requires a thick ice sheet, so AMANDA was built in the 2,800-meter-thick icecap at the Amundsen-Scott South Pole Station. The collaboration drilled holes in the ice using a hot-water drill, and lowered strings of optical sensors before the water in the hole refroze. The station provided everything from a skiway for the LC-130 turboprops that carried every piece of equipment, plus all of the fuel, to radio and internet communication, to food and housing for the summer construction crew and the two or three winter-over scientists who kept AMANDA running in the winter.

Despite the logistical difficulties, the collaboration lowered 80 photomultipliers in pressure vessels into a kilometer-deep hole during the 1993-94 season (Austral summer). Although most of the sensors survived the unexpectedly high pressures produced as the water in the hole froze, cosmic-ray muon tracks could not be seen. The problem was 50-micron-diameter air bubbles trapped in the ice, even at 1-kilometer depth. These bubbles limited the light scattering length to less than 50 cm.

Fortunately, it was predicted that near a depth of 1,400 meters, high pressure would cause the bubbles to collapse. Data confirmed this, and blue light was measured to have an incredibly long absorption length of more than 200 meters, reflecting the purity of the ice. With this understood, four strings of detectors were deployed at depths between 1,500 and 2,000 m in the 1995-96 season.

The next challenge was to separate a single upward-going muon from the roughly one million downward-going muons from cosmic-ray air showers. Algorithms with the required rejection power were developed for this, and neutrino events were identified. By 2000, the AMANDA detector was complete, with 19 strings and 677 optical sensors. A later upgrade replaced the TDC/ADC electronics with waveform digitizers. Since 2000, AMANDA-II has been recording about 1,000 neutrino events per year. For the last few years, it operated in coincidence with IceCube.

Despite this success, AMANDA's limitations were becoming obvious. It was too small, and it required manpower-intensive annual calibrations. The AMANDA optical modules (OMs) contained 8-inch photomultiplier tubes and little else; analog PMT signals were transmitted to the surface.

Although several different approaches were tried over the years, the analog transmission of signals to the surface degraded the resolution. Initially, AMANDA used coaxial cables and then twisted pairs, which transmitted high voltage for the PMT downward, and PMT signals upward. The up-to-2,500-meter cables stretched the nanosecond PMT rise time to microseconds at the surface. Although careful signal processing could provide adequate (5 to 7 ns) first-hit time resolution, there was no possibility to observe multiple hits. The long unamplified transmission required the PMTs to be run at a very high gain, near $10^9$; this had deleterious effects on the PMTs, and a few of them would occasionally 'spark,' emitting light in the process. In later strings, the electrical transmission was replaced with analog optical transmission; the PMT signal controlled an LED in the optical module. The fibers had far better time resolution, but suffered from a very limited dynamic range. The light loss in the optical couplings was very sensitive to vibration and the passage of time, so the system required manpower-intensive annual recalibrations.

These problems precluded scaling AMANDA up in size, and the collaboration began exploring other options. The most attractive, but technically challenging, option was to incorporate the digitizing electronics in each optical module. As a test, AMANDA string 18 included prototype in-OM digitizers[49]. These digitizers ran in parallel with fiber-optic analog transmission, allowing for both electronics testing and compatibility with AMANDA. String 18 performed well, and the in-OM digitization was adopted by IceCube.

**IceCube Overview**

IceCube shares many characteristics with its predecessors. As Fig.1b shows, it is a large, tracking calorimeter that measures the energy deposition in segmented volumes of Antarctic ice. Because of its size, IceCube can differentiate between electron-, muon- and tau-neutrino interactions. It has very good timing resolution, which is used to both accurately reconstruct muon trajectories and to find the vertices of contained events. IceCube is a fairly complex experiment; Table 1 lists and defines some of the IceCube-specific acronyms that are used in this review.

**Table 1.** Some IceCube acronyms and their meanings.

| Acronym | Meaning |
|---|---|
| AMANDA | Antarctic Muon and Neutrino Detection Array |
| DOM | Digital Optical Module |
| SPE | Single Photoelectron |
| ATWD | Analog Transient Waveform Digitizer |
| fADC | fast ADC |
| SCA | Switched Capacitor Array |
| SLC | Soft Local Coincidence |
| RapCal | Reciprocal Active Pulsing Calibration |

When it is completed in 2011, 5,160 digital optical modules (DOMs) will instrument 1 $km^3$ of Antarctic ice. Eighty-six vertical strings, each containing 60 DOMs, will be deployed in 2,500-meter- deep holes that were drilled in the ice by a hot-water drill. The water in the hole will refreeze, producing optical contact between PMTs and ice. The 80 strings in the baseline IceCube design will be deployed on a 125-meter grid, covering 1 $km^2$ on the surface. DOMs are attached to the strings every 17 m between 1,450 and 2,450 m. Although the minimum energy is analysis-dependent, the baseline design detects muon neutrinos down to an energy of about 100 GeV. Each string of 60 DOMs is supported by a cable that contains 30 twisted pairs (each pair is connected to two DOMs in parallel), plus a strength member and a protective covering. These cables run to a counting house in the center of the array.

Another six strings, called "DeepCore", are situated on a denser, 72-meter, triangular grid[50]. The DeepCore strings have 50 DOMs with 7-meter spacing at the bottom of the strings; 10 DOMs higher up serve as a veto. DeepCore extends IceCube sensitivity down by a factor of 10 in energy. The outer IceCube strings and top DOMs in DeepCore will serve to veto events originating outside of the central detector, greatly reducing the backgrounds for contained events. DeepCore uses newer PMTs, with higher quantum efficiency than the IceCube standard. The denser spacing and more efficient PMTs give DeepCore a lower threshold than IceCube, perhaps as low as 10 GeV.

In addition to the buried DOMs, the IceCube Observatory includes a surface air shower array known as IceTop[51]. IceTop consists of 160 ice-filled tanks, each instrumented with two IceCube DOMs. Two tanks are deployed about 10 m apart, near the top of each baseline IceCube string. Each tank is 1.8 m in diameter, and filled with ice to a depth of about 50 cm. The water is frozen in a controlled manner to minimize air bubbles. The tanks are lined with reflective material to increase light collection; depending on tank (there are small design differences as production proceeded), a typical vertical muon produces 150 to 250 observed photoelectrons[52]. The two DOMs operate at different gains, $5\times10^6$ and $10^5$, to maximize the tank dynamic range. Because of the higher data rates, each IceTop DOM has its own twisted pair.

IceTop detects cosmic-ray air showers, with a threshold of about 300 TeV. IceTop will be used to study the cosmic-ray flux and composition; the combination of air shower array data and TeV muon fluxes (observed by IceCube) provides significant handles on the cosmic-ray composition. IceTop also serves several calibration functions for IceCube. IceTop can also be used to veto high-energy cosmic-ray air showers in IceCube; conversely, one can search for muon-free showers from PeV photons.

IceCube was designed for simple deployment, calibration, and operation. Photomultiplier signals are recorded using fast waveform digitizers in each DOM. Every DOM acts autonomously, receiving power, control, and calibration signals from the surface and returning digital data packets.

IceCube construction began in the 2004-5 season, with the first string deployment. By January 2010, 79 of the total (including DeepCore) strings had been deployed, and the array should be complete by January 2011.

### IceCube Construction and Operations

Construction at the South Pole is difficult, and logistics are tough. The construction season is short, from mid-October through mid-February, and being able to drill holes and deploy strings quickly is critical. In order to be able to build IceCube in seven seasons, holes had to be drilled in less than 2 days, requiring a power plant of close to 5 megawatts to melt ice. Specialized equipment was designed for this effort.

Getting the roughly 1 million pounds of drilling equipment to the South Pole was another major challenge. The drilling equipment had to be built in modular form, with each component able to fit into a LC-130 transport plane. Because of the high altitude, the plane's payload is limited, further straining the logistics chain.

IceCube DOMs are deployed in water-filled holes, 61 cm in diameter. The water at the edges of these holes begins to refreeze almost immediately; their 61-cm diameter insures that the holes remain open wide enough to accommodate the cable and DOMs for 30 hours; this allows a full string deployment.

The ice is melted by a hot water jet under pressure from a nozzle supplying 200 gallons per minute at 6.89 MPa and a temperature of 88 C. The water is subsequently pumped out of the hole and returned to a heating system at a rate of 927 l/minute and a temperature of 1 C. The water is reheated at the surface and returned to the nozzle at depth. Drilling progresses by circulating this water, reheated at the surface, to ever-increasing depth. Also, 30 l/minute is added to the loop at 1 C to compensate for volume reduction when the ice becomes water. The operation requires a heating plant generating 4.7 MW in thermal and 300 kW in electrical energy. It can drill holes in less than 30 hours[14], using less than 30,000 liters of fuel per hole. Figure 7 shows a photo of the drill head in operation.

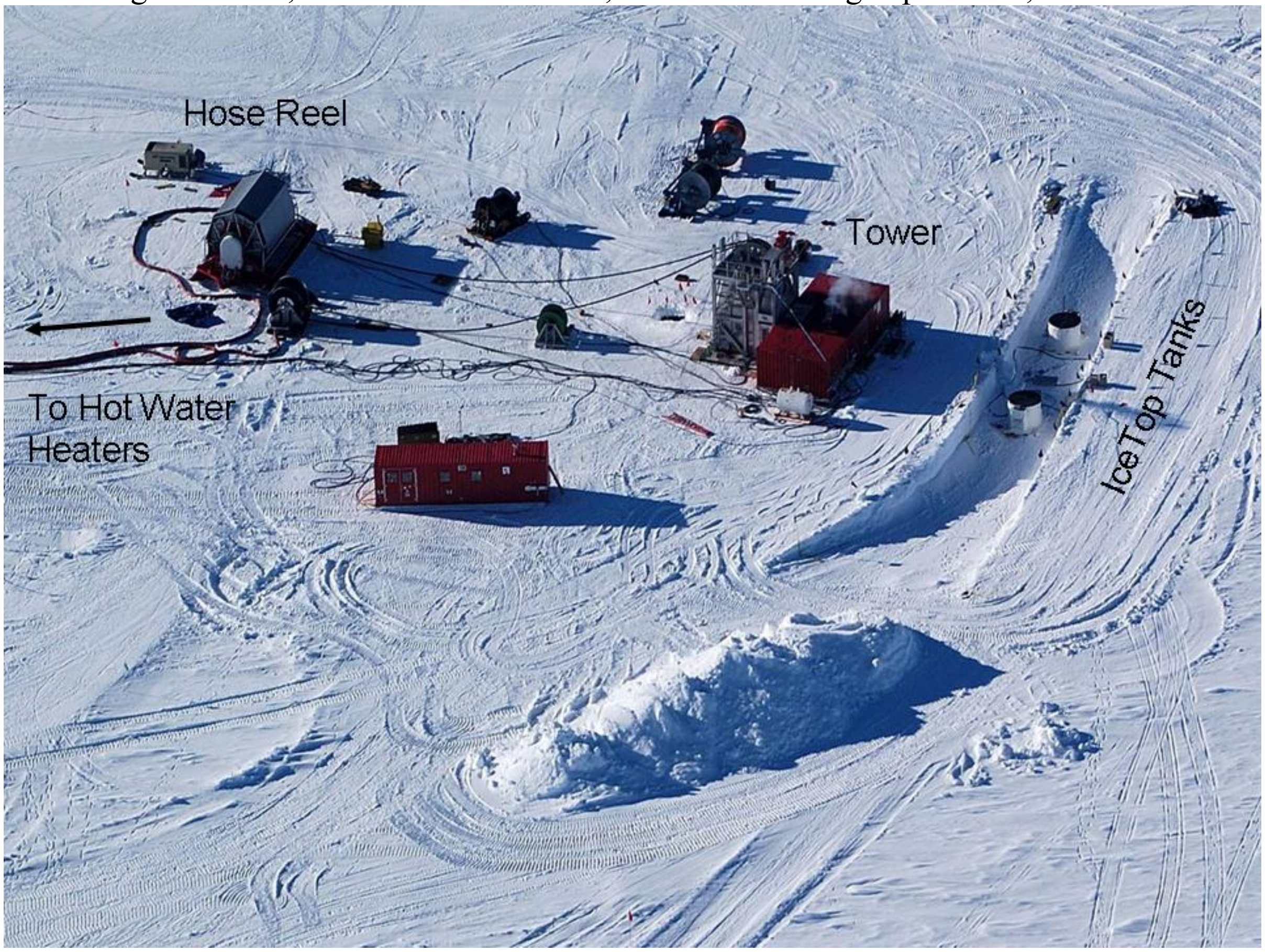

**Fig.7** IceCube drilling site at Amundsen-Scott South Pole Station. The hole into which the optical sensors will be lowered is drilled under the tower building to the center-right. Hot water is pumped from the heaters (not shown) producing 5 MW of hot water under 1000 psi pressure through the hose at the left. The cylindrical hose reel holds 2500 meters of hose which unreels as drilling proceeds. The trench at the right holds two IceTop tanks filled with ice. From Ref. 2.

The system has many components. This summary traces the chain of operations that delivers a hole filled with water of sufficient width and depth to deploy a string of optical sensors:

- The firn drill: an independent, 150 kW, electrically heated glycol loop powered by a generator that melts holes through the top layer of about 50 m of snow that is porous to water.
- The Rodriguez well: formed each season by operating a hot-water drill at a fixed depth, thus creating a cavity of water. This water is used to initiate drilling and to supply the replacement water previously mentioned.
- The preheat system: 4 car-wash-style heaters and 12 “stinger” heaters bring water sequentially to 10 C and 21 C in two 38,000-liter tanks.
- The main heating plant: consists of 35 high-efficiency, car-wash-style heaters that deliver water to the drill head at 85 to 66 C, depending on depth.
- High-pressure pumps: 4 units provide 760 l/minute flow under a nominal pressure of 6.89 MPa.
- The drill supply-hose reel contains 2,700 m of hose on a motorized reel with level wind and brake to supply hot water to the drill head. The hose inner and outer diameters are 10 and 15 cm, respectively; see Fig.7.
- Two towers: they leapfrog from hole to hole, changing the supply hose from horizontal to vertical.
- Drill head: exit nozzle with a weight stack to maintain a vertical hole by gravity. Water exits the nozzle at a speed of 48 m/sec, achieving a drill speed of over 2 m/min.
- Return-water hose reel for the 50 Hp return-water pump.
- A closed-loop computer control system with more than 400 input/output points.
- The pump returns the water from the hole to the first surface tank at 1 C for reheating to 88 C, thus closing the loop.

The fuel tanks and all drill components are built on movable sleds to allow for repositioning the drilling infrastructure every new season. The drill speed is computer-controlled; the actual width and refreeze time of each hole are accurately computed on the basis of drill data entered into the software. In the 2009-10 season, the system delivered 20 holes with a performance far superior to design. Holes were drilled in as little as 27 hours, with a fuel consumption of just over 15,000 l, greatly improving upon design goals.

Before the water in the hole refreezes, a cable is lowered into the hole that will carry the signal to the surface from the 60 DOMs attached every 17 m. In-situ construction of the string and lowering it to depth takes roughly 10 hours. Each string takes data as soon as the hole refreezes.

## The Ice in IceCube

The ice surrounding the DOMs serves as a Cherenkov radiator. Optical absorption and scattering of the radiated photons are both important in determining what IceCube observes. The optical transmission depends strongly on impurities in the ice. These impurities were introduced when the ice was first laid down as snow. This happens in layers; each year snowfall produced a thin, nearly horizontal layer. For the ice in IceCube, this happened over roughly the last 100,000 years. Variations in the long-term dust level in the atmosphere during this period, as well as the occasional volcanic eruption, lead to depth-dependent variations in the absorption and scattering lengths.

Because the bulk of the scattering is in the forward region, light scattering in ice is usually parameterized in terms of the effective scattering length,

$$\Lambda_{eff} = \frac{\Lambda_{scat.}}{1 - \langle \cos(\theta) \rangle} \qquad (11)$$

where θ is the mean scattering angle per scatter. $\Lambda_{eff}$ is, of course, frequency dependent.

Much effort has gone into measuring the optical properties of the ice, using artificial light sources and in-situ measurements. In AMANDA and IceCube, studies have been done using LEDs and lasers that emit at a variety of wavelengths. The AMANDA data is still valuable because it involves measurements at many wavelengths. By measuring the arrival time distributions of photons at different distances from a light source, it is possible to measure both the attenuation length and scattering length of the light. These measurements, although useful, suffer from a limited resolution in depth[53].

Higher-resolution depth-dependence measurements of the ice properties come from 'dust loggers' which are lowered down water-filled holes immediately after drilling. They shine a thin beam of light into the ice, and measure the reflected light[54]. This can measure the ice properties with a vertical resolution given by the width of the emitted beam – a few mm. Figure 8 compares some of the dust logger data with optical scattering measurements and with ice cores collected elsewhere in Antarctica.

Fig.9 shows the absorption and scattering distances currently used in IceCube as a function of depth and wavelength. These curves are based on theoretical models fit to the AMANDA and IceCube measurements. The effect of air bubbles at shallower depths is clearly visible, along with broad dust peaks and the native absorption in the ice. Not visible are the very narrow peaks due to thin layers of dust produced by volcanoes.

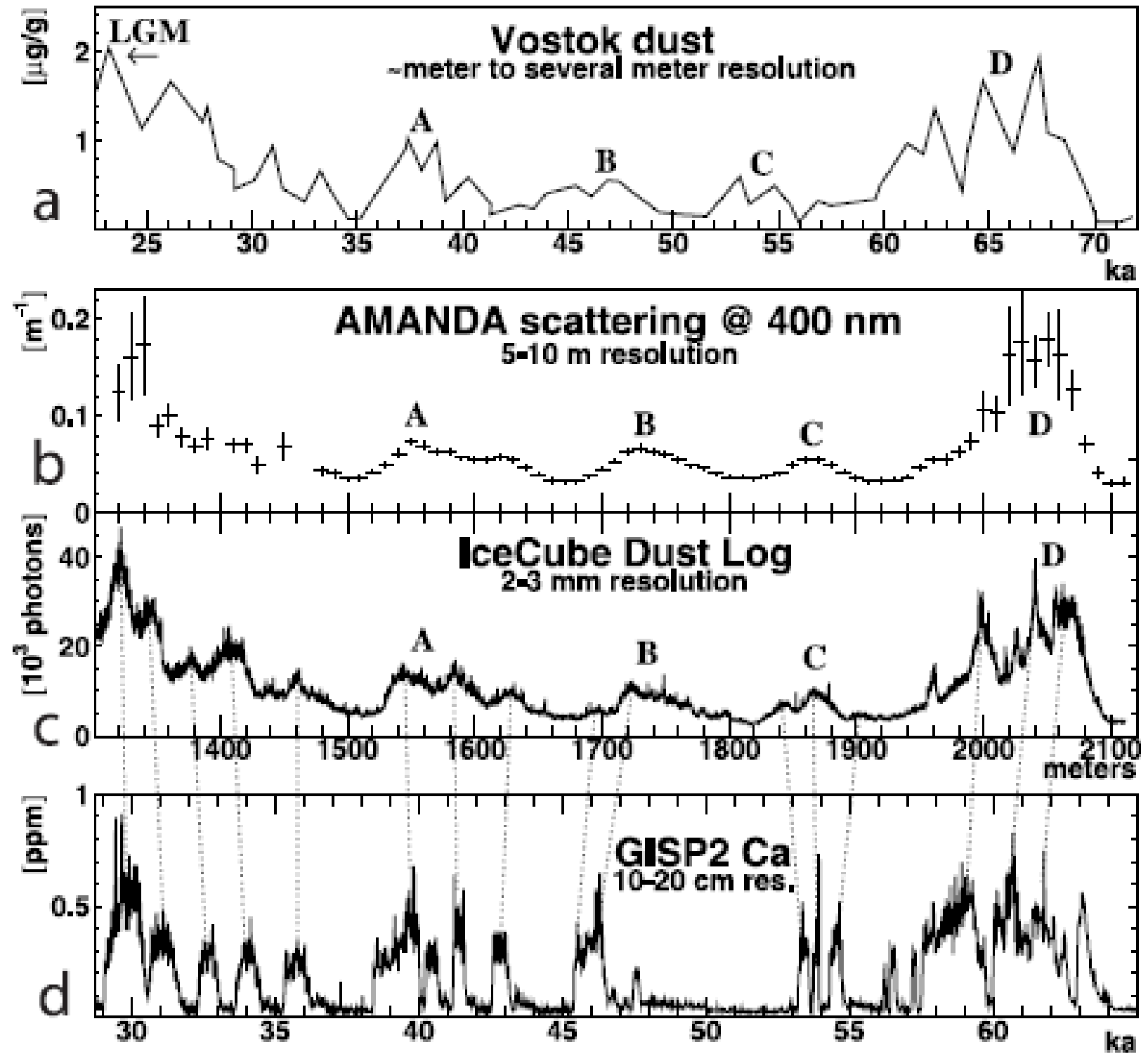

**Fig. 8**. Data from the IceCube dust logger, compared with AMANDA measurements based on light scattering from sources to optical modules, along with measurements from two other Antarctic sites[54].

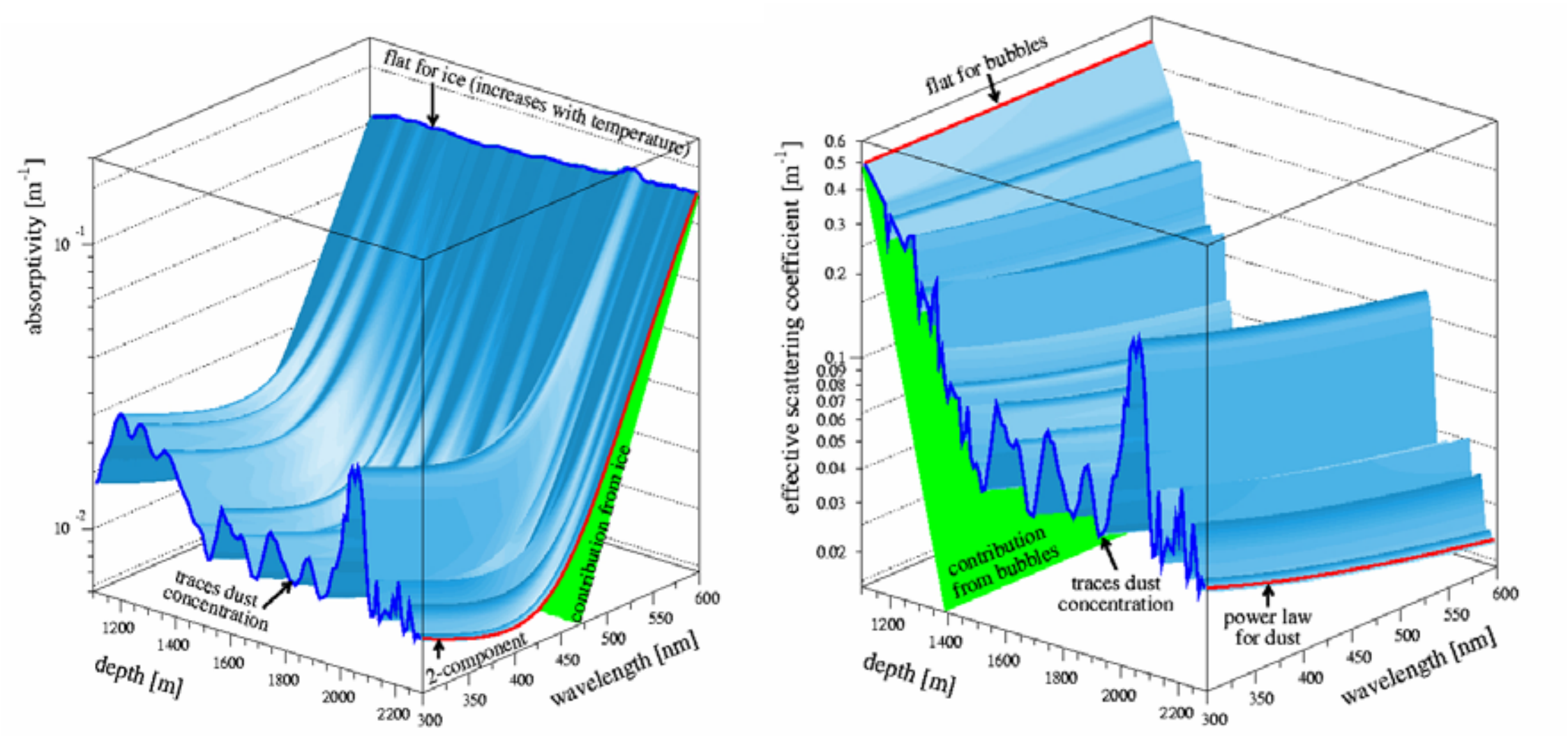

**Fig.9** Absorption (left) and scattering (right) lengths of light in South Polar ice, as a function of depth and wavelength, from 300 to 600 nm[53].

## Digital Optical Module Hardware

The DOMs had rather stringent design requirements[11, 55]. They had to record the arrival times of most of the photoelectrons observed by the photomultiplier tube, with a timing resolution (across the array + IceTop, a spread of up to 3 km) of less than 5 ns. The dynamic range requirement was 200 photoelectrons per 15 ns. The PMTs also had to have a dark noise rate less than 500 Hz; this in turn set limits on the radioactivity in the pressure vessel and on the PMT. DOMs had to work from room temperature (for testing) down to -55 C. Because of the high cost of power -- fuel for the generator must be flown in on LC-130's - each DOM was required to draw less than 5 Watts.

Finally, because the DOMs are totally inaccessible after deployment, we set a reliability requirement of 90% DOM survival after 15 years. This is the same reliability as is expected for satellites, but on a much lower budget.

Fig.10 is a schematic diagram of a digital optical module. Each DOM consists of a 10 in (25 cm) photomultiplier tube and associated electronics, plus a data acquisition system, all in a 35-cm-diameter pressure sphere. The PMT electronics include a Cockroft-Walton high-voltage power supply and resistive divider. The DOMs also contain 13 light-emitting diodes used for photonic calibrations.

The pressure vessel is a 0.5 in-thick borosilicate glass sphere capable of withstanding a pressure of 70 MPa. The glass transmits light with a wavelength longer than about 350 nm, limiting the high-frequency response of the DOM. Radioactive decays in the glass are a significant contributor to the PMT dark noise; the resulting fluorescence in the glass produces time-correlated background in the PMT, out to times of a few μs. The PMTs are optically coupled to the pressure vessel using optical coupling gel. The sphere is filled with nitrogen gas at ½ atmosphere pressure.

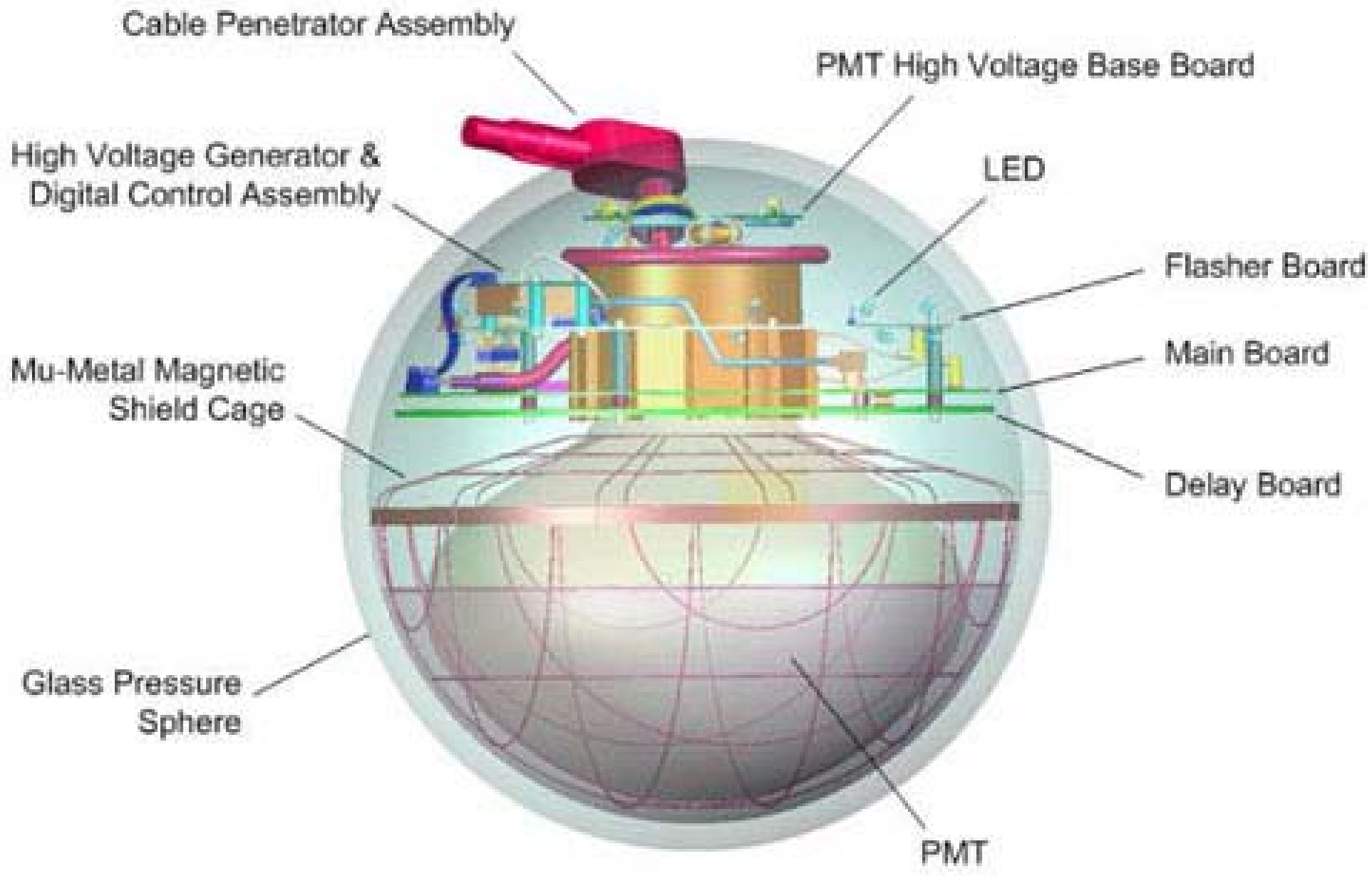

**Fig.10** Schematic drawing of a digital optical module.

**The Photomultiplier and Associated Circuitry**

The baseline IceCube and IceTop DOMs use Hamamatsu R7081-02 photomultiplier tubes[56]. These tubes have standard bialkali (Sb-Rb-Cs and Sb-K-Cs) photocathodes, sensitive to 300-650 nm photons, with a 25% peak quantum efficiency. The amplifying section has 10 linear-focused dynodes, and runs at a gain of $10^7$ at a nominal 1500 V. Most of the PMTs are run at a gain of $10^7$. The typical high voltages range from 1300 to 1500 volts. A mu-metal shield surrounds the PMT and reduces the ambient (Earth's) magnetic field in the PMT by about a factor of two.

The PMT bases are conventional resistive dividers, with the Hamamatsu-recommended ratios and a total resistance of 130 MΩ. High resistances were used to minimize power consumption. Capacitors are placed across the last 6 dynode stages to maintain the voltages in the presence of large pulses; their recharging time-constants are of order 1 second. The capacitors are sized so that the PMT loses less than 20% gain for a $10^6$ photoelectron pulse.

The high voltage is supplied by custom-designed[57] Cockroft-Walton power supply. It is both low-power (<300 mW) and low-noise (< 1 mV). The output voltage is digitally controlled, and may be adjusted from the surface.

The PMTs operate with the cathode grounded; the anodes are at a high potential. The anode signals are coupled to the data acquisition electronics with a bifilar wound toroidal transformer. A transformer was used instead of a capacitor because it has much lower stored energy for an equivalent frequency response. The transformers were designed to have a frequency response from 8 kHz to over 100MHz, down to –40 C. A square wave input produces an output signal with a decay time of more than 15 μs, far longer than the lengths of a single event (except for possible slow magnetic monoples). The first 1,200 DOMs were built with an earlier design, with a 1.5 μs time constant. During data analysis, this droop is removed with a digital filter. The only problem occurs when the ADCs overflow or "bottom out", in which case some information is lost.

The system response to a single photoelectron (SPE) is a pulse with roughly triangular waveforms, with average amplitude of about 10 mV and a width of 5 ns.

Fig.11 shows the charge spectrum produced by the PMT running at a gain of $10^7$ in response to a low-amplitude LED (emitting much less than 1 photoelectron/pulse). The single photoelectron peak is described by a Gaussian with a resolution of about 30%. However, about 15% of the SPE produce a much smaller output, less than 0.3 of the peak charge. This low-charge tail is from real SPE.

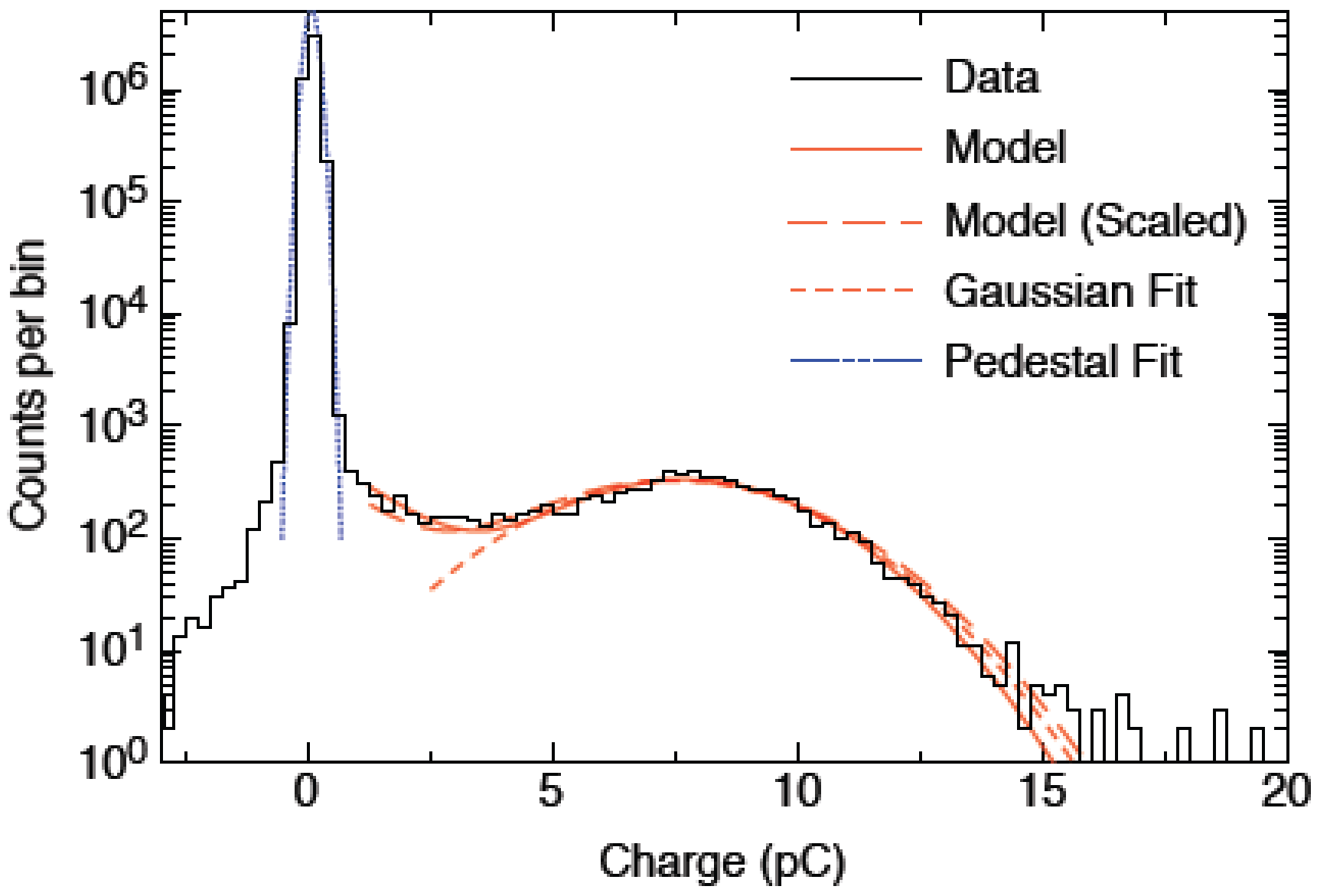

**Fig.11** The single photoelectron charge spectrum observed in the PMT at a gain of $10^7$. From Ref. 56.

Fig.12 shows the arrival times of single photoelectron pulses. For single photoelectrons, the time resolution is about 2 ns, although a tail of late-arriving pulses is clearly visible, with peaks around 30 ns, 75 ns and 130 ns late; about 4% of the pulses come more than 25 ns after the expected arrival time. The timing is slightly sensitive to where the photoelectrons hit the photocathode; photons striking the edges of the PMT are recorded, on average, about 3 ns later than those reaching the center. Their times resolution is also worse.

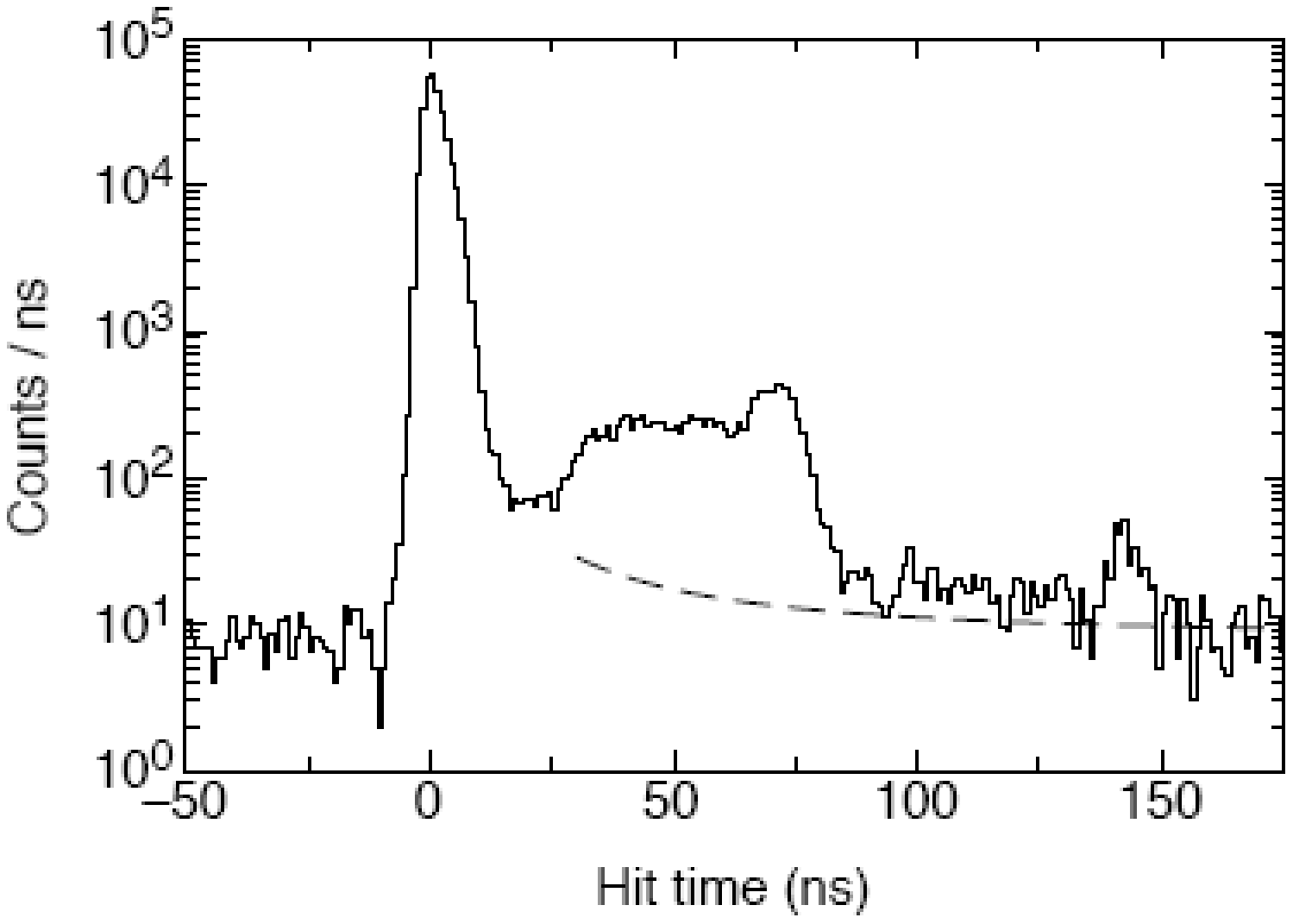

**Fig.12** The distribution of arrival times of single photoelectrons at the PMT. A tail of late-arriving photons follows the main pulse[56]. The dashed line shows the contribution to late light due to laser afterglow plus random background.

Since nearby neutrino interactions can produce large signals in the IceCube DOMs, the PMT response to large signals was well characterized. Fig.13 shows the signals produced by light pulses containing about 220, 3,700 and 210,000 photoelectrons. Saturation is clearly visible for the two larger signals, and the PMT output is no longer linear.

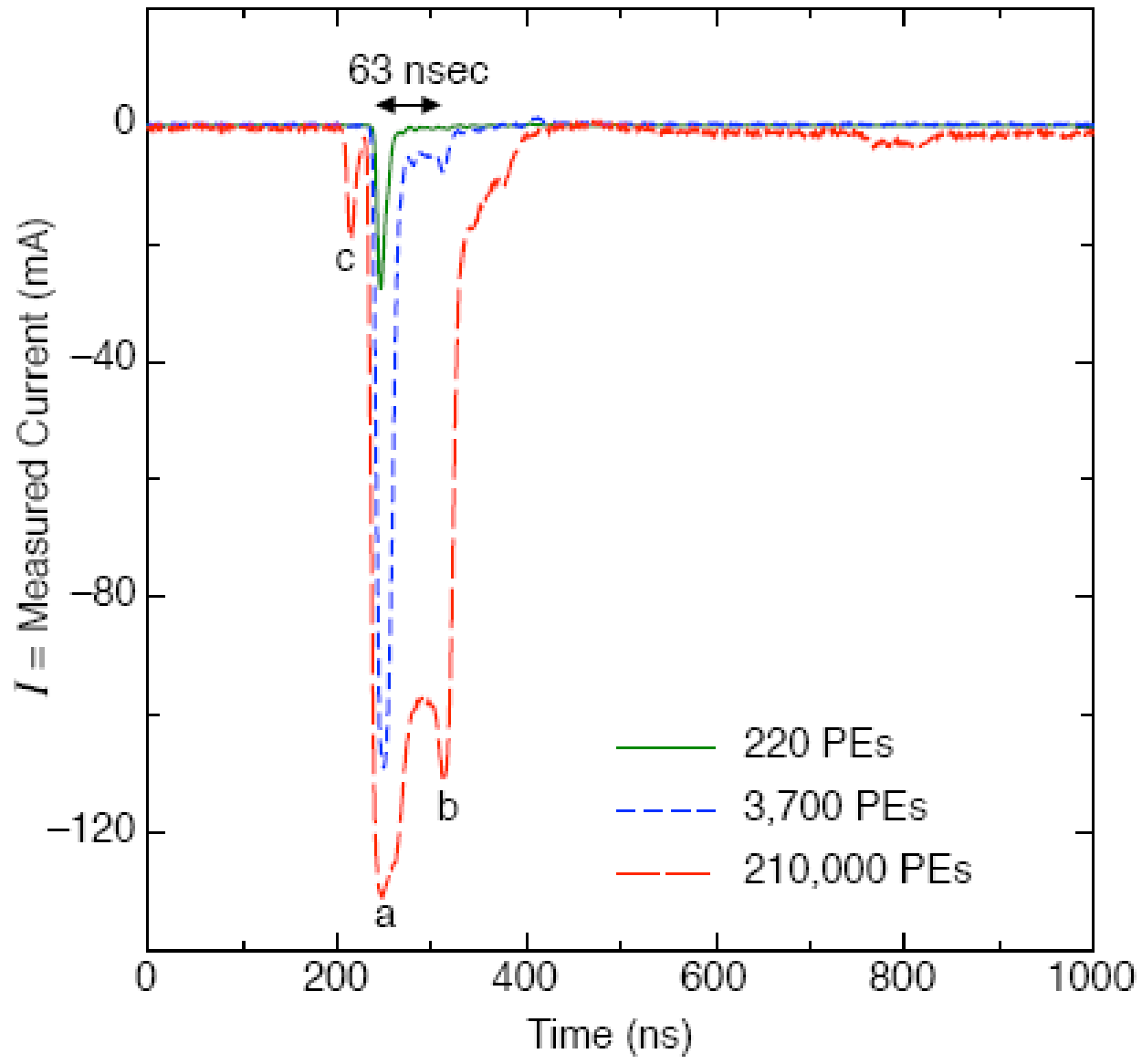

**Fig.13** PMT signals for different input amplitudes, showing the effect of saturation. Saturation compresses the instantaneous current, so that small 'features' like prepulsing and afterpulsing grow in relative size when saturation sets in[56].

The saturation appears to be an instantaneous effect, and can be corrected solely modifying the PMT current, without reference to the previous history of the pulse. The corrected current $I_0$ is found from the measured current $I$ using

$$I_0 = \ln(I) + \frac{C\left(\frac{I}{A}\right)^{B}}{\left(1-\frac{I}{A}\right)^{1/4}}$$

where A, B and C are constants that vary significantly from PMT to PMT; they also depend on PMT gain.

## DOM Data Acquisition Electronics

Fig.14 shows a block diagram of the DAQ system, and Fig.15 shows a photo of the main board[58]. The central elements of the DAQ hardware are two waveform digitization systems, the Analog Transient Waveform Digitizer (ATWD) and the fADC ('fast' ADC).

A digitization cycle is initiated by a discriminator trigger; the voltage threshold corresponds to about 1/4 photoelectron. When the discriminator fires, the FPGA starts ATWD and fADC digitization synchronously, on the next clock edge.

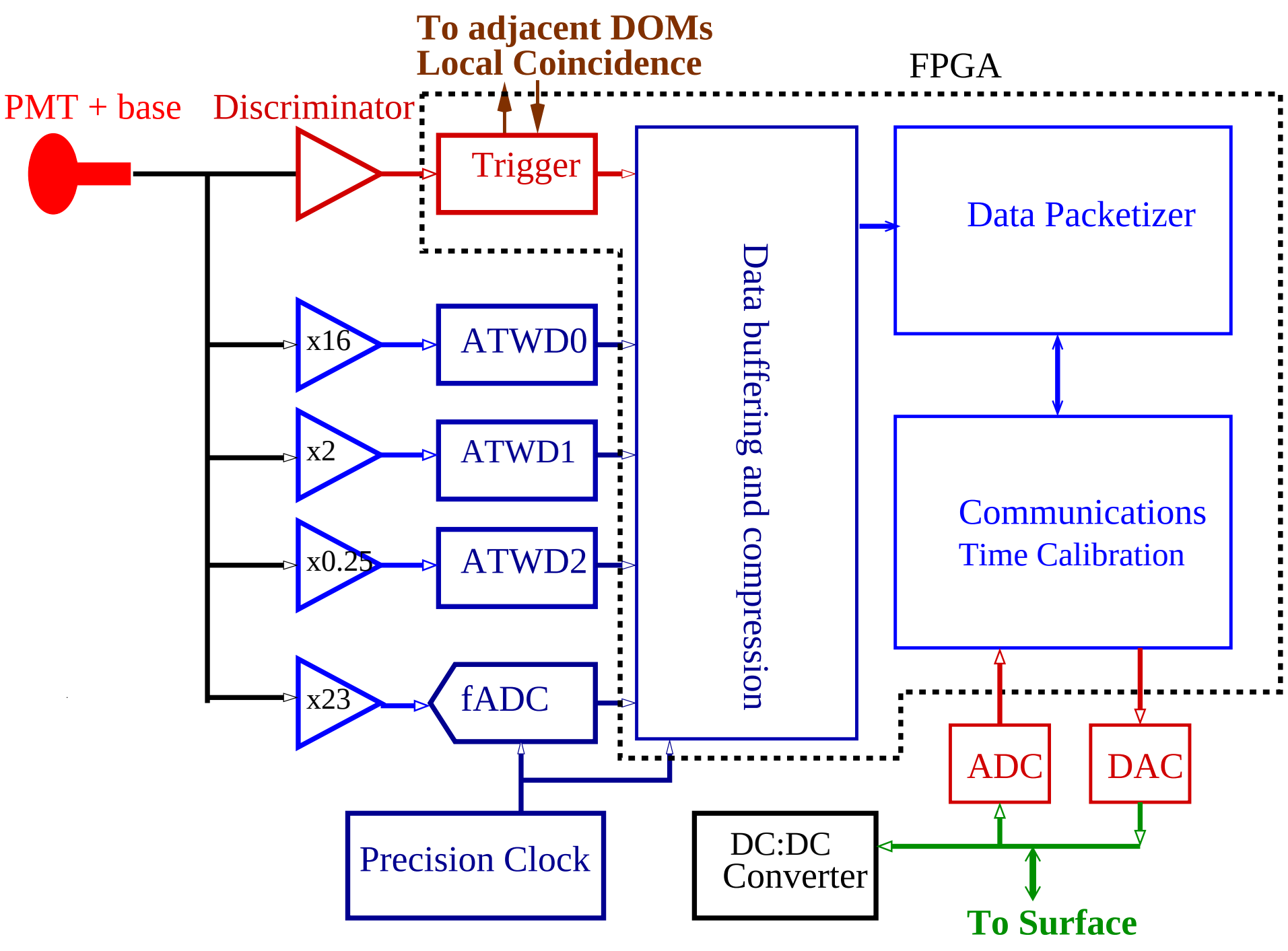

**Fig.14** A simplified block diagram of the IceCube main board electronics[59].

A 70-ns-long delay line is used to delay the signal to the ATWDs so that the PMT pulse appears near the beginning of the digitization cycle. To maximize reliability, this delay line is implemented on a separate printed circuit board containing a long, winding trace. This approach limits the delay line bandwidth to about 100 MHz. The amplifiers that feed the ATWD chips also have about 100 MHz of bandwidth.

The ATWD digitizer is a custom CMOS analog integrated circuit containing a switched-capacitor array (SCA)[60]. Each channel contains an array of 128 capacitors connected to the input via a set of switches, which are normally open. When an acquisition cycle is launched, the switch to each capacitor is closed, in turn. The switches are controlled by an adjustable delay line. By varying a control voltage, sampling rates between 200 and 900 MegaSamples per second (MSPS) are possible; IceCube runs the ATWDs at 300 MSPS. The ATWD chip consumes about 30 mW, far less than any comparable commercial digitizer.

Each ATWD chip has four parallel inputs. Three ATWD channels connect to the PMT signal, with input gains in the ratio of 16:2:1/4, providing more than 14 bits of dynamic range. After acquisition, the voltages on the capacitors are digitized with 128 10-bit Wilkinson ADCs, each multiplexed to the four capacitors, which acquire a single time sample. The fourth ATWD input (not shown) is used for electronics calibrations. Digitization takes 29 µs per waveform.

The analog performance of the chips is ample for IceCube. The analog bandwidth of the sampling circuitry is higher than 350 MHz, higher than the circuitry that precedes it. The analog dynamic range of 2500:1 is higher than for the 10-bit outputs. However, some effort is required to calibrate the data. The relationship between control voltage and sampling rate varies from chip-to-chip. In IceCube, the 4th ATWD channel digitizes the 40-MHz clock; from this the sampling rate is determined. Each capacitor has its own pedestal value; the 512 pedestals from a single chip must be determined individually and stored.

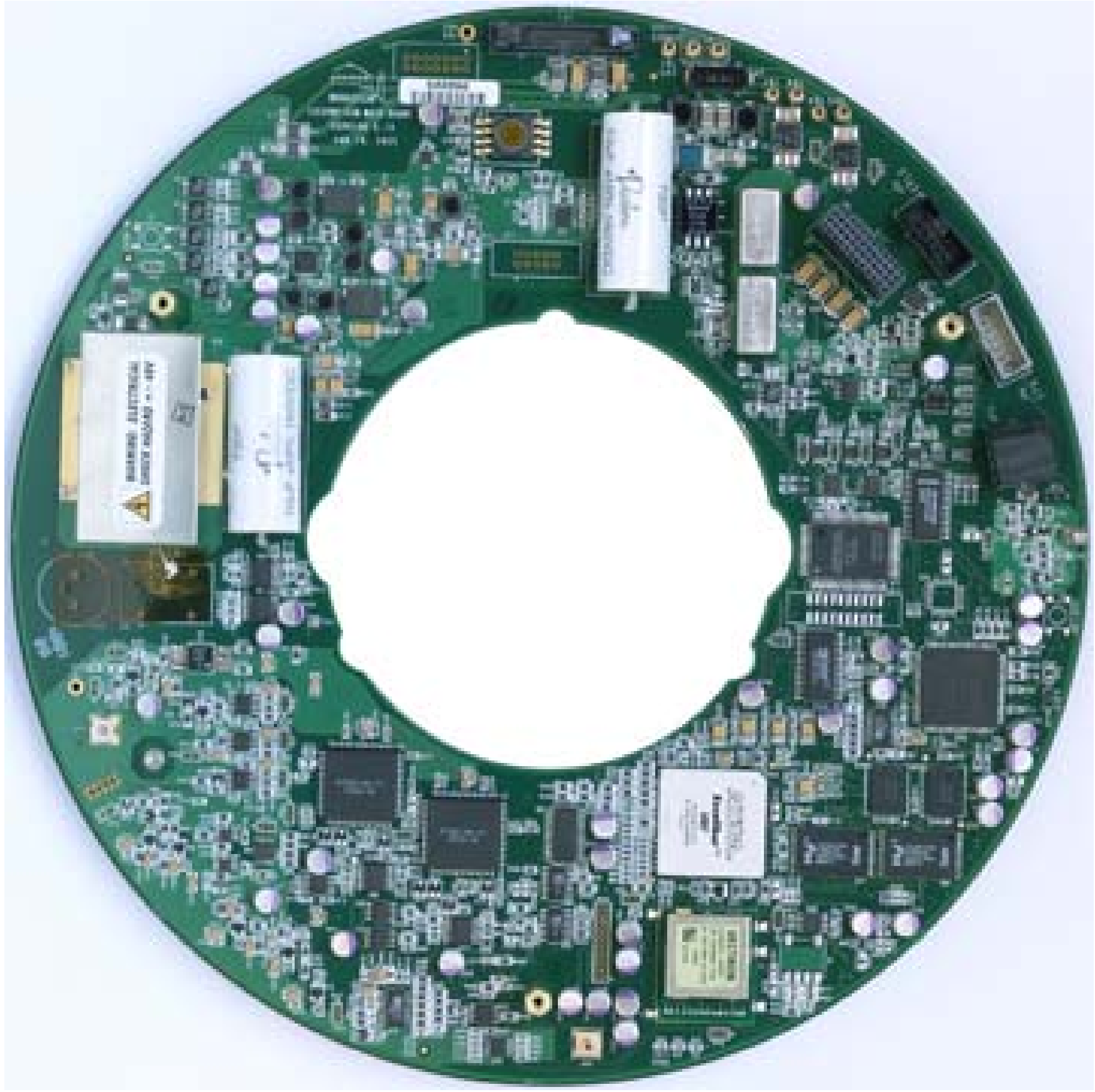

**Fig.15** A photograph of the DOM Main board. The circular board fits in the pressure vessel, while the cutout provides room for the neck of the PMT[58].

Each DOM contains two ATWD chips. They operate in ping-pong fashion – while one is digitizing, the other is live; this greatly reduces the dead time.

The fADC digitizer system uses a continuously running, commercial 10-bit 40 MSPS digitizer chip which runs continuously. It is preceded by a 3-stage shaping amplifier, which lengthens the PMT pulses with a 180-ns shaping time, to better match the sample time. Since the pulse covers multiple samples, it is possible to determine the photon arrival time to better than 5 ns. When a trigger occurs, the system records 256 fADC samples, covering 6.4 μs. The fADC suffers from limited dynamic range; it overflows for even medium-sized signals. However, for smaller signals, it can provide important information about late-arriving light.

The data from a single trigger ("launch" in IceCube parlance) consists of at least one ATWD waveform and one fADC waveform, plus a time stamp and the local coincidence signals from the adjacent DOMs. The highest-gain ATWD waveform data is always saved. If any single bin of that waveform is above 768 ADC counts, then the medium-gain channel is also read out. If any bin of the medium-gain channel is above 768 counts, then the low-gain channel is also saved. To reduce bandwidth, the waveforms are compressed using “delta-compression”. Each sample (except the first) is replaced with the difference from the previous sample; these are mostly small numbers. Then, these deltas are stored using variable width symbols.

The time stamp comes from the 40-MHz system clock, showing when the DOM “launched.” Since this determines the arrival time of every photon in the event, accuracy is crucial, and a precision oscillator is used. Power limitations precluded the use of an oven-controlled oscillator, but, fortunately, the temperature is very stable. The system uses an oscillator with a frequency stability (Allen variance) of better than $\delta f/f < 10^{-10}$, adequate to maintain the required 5-ns precision over tens of seconds; this sets the required interval between timing calibrations, as is discussed below.

The entire system is controlled by a 400k-gate Altera Excalibur EPXA-4 FPGA, which incorporates an ARM9 hard-core CPU. The FPGA controls the trigger and digitizer, buffers and zero-suppresses the data, and does most of the packet assembly, while the CPU performs higher-level tasks, including some calibration work.

One challenge for this system was to ensure that the firmware and software were upgradable, while at the same time making sure that a bad ‘load’ could not cause a permanent loss of communication. For this, the FPGA uses two programming sources. On power-on, the FPGA boots from a one-time programmable 8-Mbit configuration memory that provides basic functionality such as communications and a simple utility program. This memory cannot be altered. However, to allow for reprogramming, it can then be directed to switch its programming, using data from an 8-Mbit flash memory which can be reprogrammed from the surface.

A few functions which cannot be implemented in the FPGA are performed by a CPLD. This CPLD retains its logic configuration even without power. It controls the FPGA configuration cycle and the interface to the flasher board, reads 24 ADC channels used for monitoring, and controls 16 DAC channels that provide analog control voltages. Fig.16 shows the digital interconnections on the main board.

The FPGA includes a 4-bit scalar, which counts the number of PMT pulses in each 1.6-ms period. The 640 scalar readings/second are stored in memory and read out periodically. They are used to search for bursts of neutrinos due to a supernova, as is discussed in Section V.

Data from two DOMs is transmitted to the surface via a single twisted pair, which also provides ±48 VDC (96 volts total) power. Each DOM consumes about 3.5 W. The signals are transmitted using an 8-bit DAC, and received with a 10-bit ADC; both run at 40 MSPS. This is overkill for the 900 kbit/s data communication rate using amplitude-shift modulation, but is important for the RapCal timing calibration described below. Higher rates would be possible using more sophisticated protocols, but they are not needed.

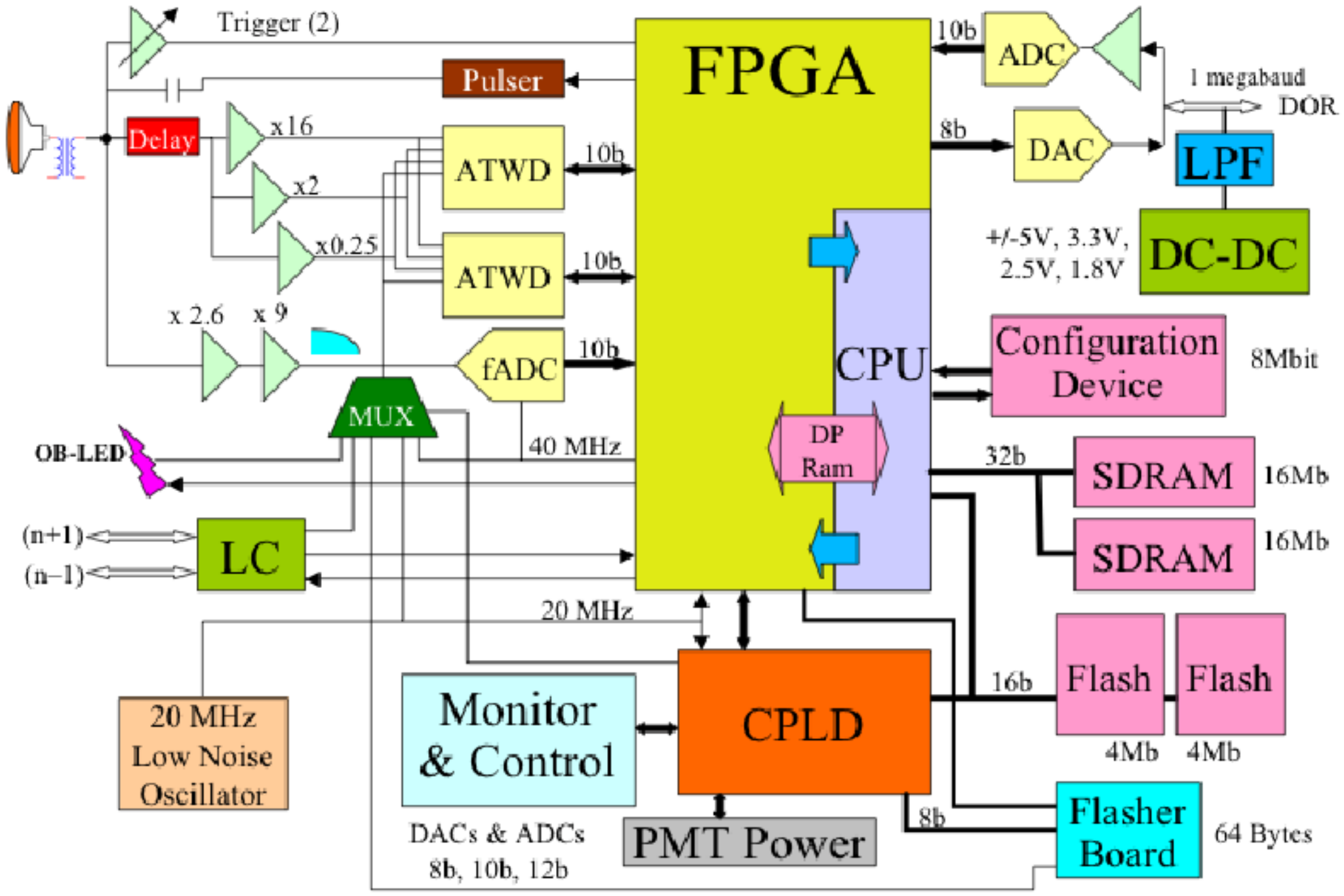

**Fig.16** Diagram of the digital electronics, including the calibration and monitoring circuitry.

The cable incorporates local coincidence circuitry, whereby DOMs communicate with their nearest neighbors; they can also pass messages onward. IceCube DOMs have several operating modes for the local coincidence circuitry. Until early 2009, IceCube ran in "Hard Local Coincidence (HLC)" mode. In this mode, the DOMs only saved data when two nearest neighbor or next-to-nearest-neighbor DOMs saw a signal within a 1-µs coincidence window. When this happened, the entire waveform information was sent to the surface. The HLC hit rate depends on a DOM's depth, through both the muon flux and the optical properties of the ice, but is typically 3 to15 Hz; the noise rate for HLC hits is very low.

In early 2009, IceCube started taking data in "Soft Local Coincidence (SLC)" mode. In addition to the complete waveform data for coincident hits, a "Coarse Charge Stamp" was sent to the surface for isolated hits. These charge stamps contains three fADC samples; the highest 3 samples out of the first 16, along with the time of the highest sample. SLC hits are recorded at the PMT dark rate, typically 350 Hz. Although most of these hits are noise, they are useful in many analyses, especially when a preliminary reconstruction can be used to restrict the active time and volume for SLC hits.

One other critical requirement for IceCube hardware is reliability. Once deployed, it is impossible to repair a DOM, so the system demanded very high reliability. Several measures were taken to assure this. High-reliability parts were used where possible, and all were heavily derated. The PC board was laid out with conservative design rules, and was constructed from high-temperature FR4 (chosen because of its lower-temperature coefficient). Finally, the boards and completed DOMs were subject to stringent testing. Prototype boards were subjected to HALT (Highly Accelerated Lifetime Test) cycling, including high and low temperatures, rapid temperature cycling, and high vibration levels. Thermal imaging was also used to check for hot spots. All of the production boards were subjected to HASS testing, a less-stressful version of HALT. The testing culminated in a month-long burn-in of the complete DOMs, which included cycling from room temperature to –55 C. Ninety-eight percent of the DOMs survive deployment and freeze-in completely; another 1% are impaired, but usable (usually, they have lost their local coincidence connections). Post-freeze-in, reliability has been excellent, and the estimated 15-year survival probability is 94%.

## Hardware Calibrations

Determining the time and amplitude of an observed light pulse requires good calibrations. IceCube uses a variety of methods to ensure good calibrations.

The primary timing calibration is "RapCal": Reciprocal Active Pulsing[61]. RapCal timing calibrations are performed automatically every few seconds. During each calibration, the surface electronics send a timing signal down to each DOM, which waits a few μs until cable reflections die out, and then sends an identical signal to the surface. The surface and DOM electronics use identical DACs and ADCs to send and receive signals, so the transmission times in each direction are identical. Fig.17 shows typical waveforms after 3.5 km of cable. Even though the 3.5-km cable transmission widens the signals to ~1 μs, the transmission time is determined to less than 3 ns[62].

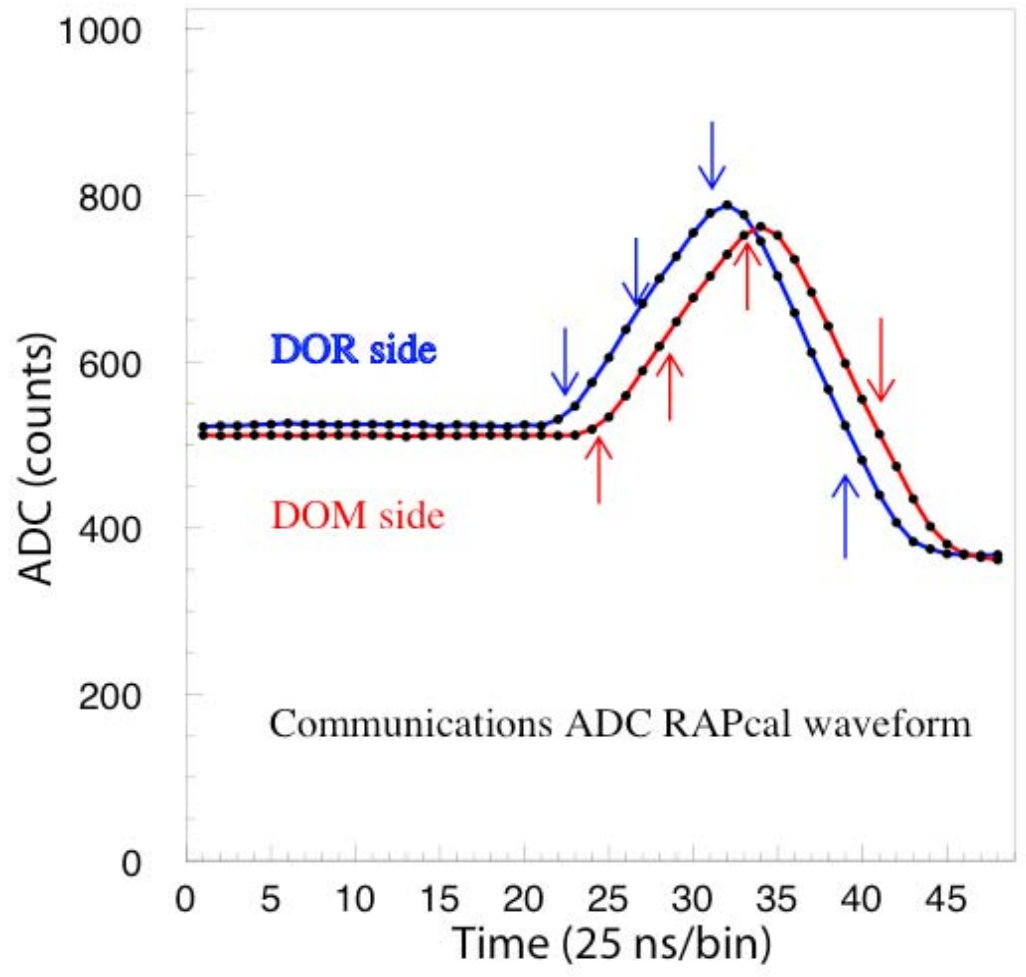

**Fig.17** RapCal timing waveforms, as received by the DOM, and on the surface ("DOR side"). Initially narrow pulses are now ~ 1 μs wide[62].

Other timing calibrations measure the signal propagation delay through the PMT and electronics. Each main board includes a UV LED ("On-Board LED" in Fig.16), which may be pulsed on command. The LED pulse current is recorded in the ATWD, along with the PMT signals. The difference determines the PMT transit time, plus the delay in the delay line and other electronics.

Amplitude calibrations are also done with the On-Board LED. It is flashed repeatedly at low intensity (<< 1 photoelectron in the PMT). A charge histogram is accumulated in the FPGA and sent to the surface, where it is fit to find the single photoelectron peak. This is done for a range of high voltages, and the high voltage is set to give $10^7$ PMT gain. These calibrations are extremely stable over time periods of months.

Each DOM also contains a 'flasher' board with 12 LEDs mounted around its edges. These LEDs are used for a variety of calibrations, measuring light transmission and timing between different DOMs[62]. The multiplicity of LEDs is particularly useful for linearity calibrations. The LEDs are flashed individually, and then together, providing a ladder of light amplitudes that can be used to map out the saturation curve.

Calibrations are also studied using cosmic-ray muons, plus two special devices – the "Standard Candles"[63]. These are extra modules containing a 337-nm $N_2$ laser mounted between two DOMs on a cable. The laser beam is shaped to emit light in the shape of a Cherenkov cone, forming a reasonable approximation to a cascade. The light output is well-calibrated, and an absorber wheel allows for variable intensities. Although the 337-nm light does not propagate as far as typical Cherenkov radiation (peaked around 400 nm, after factoring in detection probability), it provides a reasonable simulation of cascades up to PeV energies.

Figure 18 shows one of the higher-level time calibrations, using the LED flashers. The time difference is that expected for the DOM-to-DOM separation, and the RMS times for all of the adjacent DOM pairs on the string are between 1 and 2 ns. It is worth pointing out that, although the DOMs are adjacent, the signals only come together at the surface, so there is effectively 5 km of cable separating the two DOMs. Other studies, using muons, give similar timing resolutions; the relative timing calibrations are stable over time periods of at least months.

## Surface Hardware, Triggering and Filtering

Figure 19 shows a block diagram of the surface electronics. Almost all of the hardware is commercial; the only exceptions are the "DOR cards" which receive the signals from 8 DOMs. The cards plug into standard PCI slots in standard industrial PCs called String Hubs; Fig. 20 shows a block diagram of a String Hub. Each String Hub holds 8 DOR cards, so one hub can control an entire detector string. Because of the higher data rate, the IceTop DOMs are plugged into a separate set of hubs, with 32 DOMs per hub.

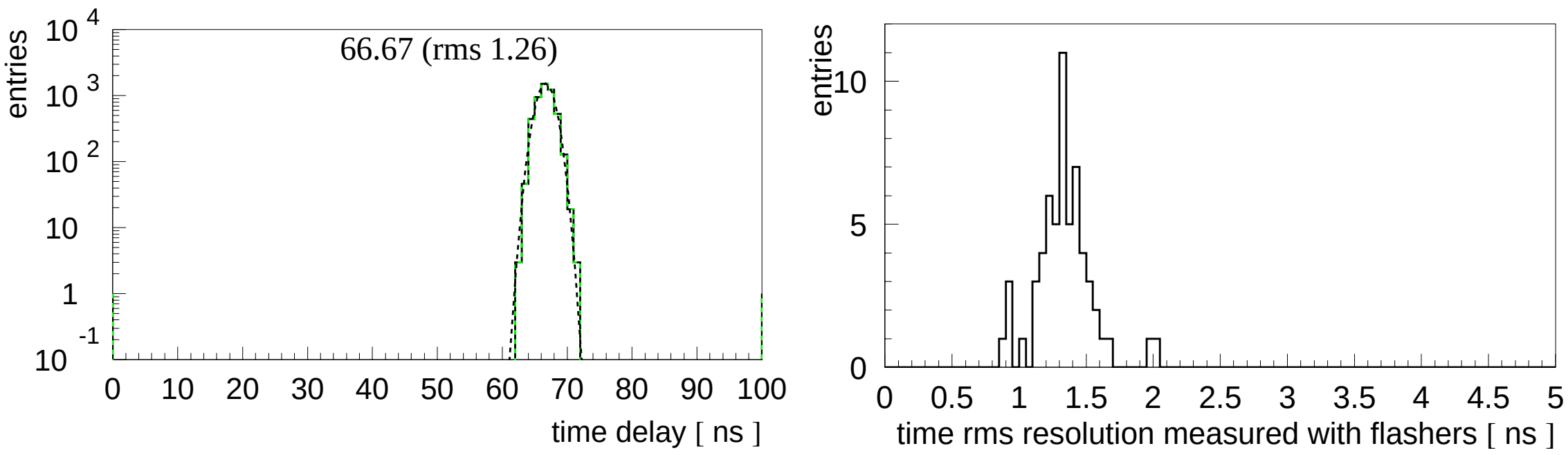

**Fig. 18** (Left) The time distribution of the first photons arriving at DOM 46, String 21, when DOM 47 on the same string is flashing; the time difference is consistent with the 17 m separation, and the 1.26 ns sigma shows that the relative timing is accurately calibrated. (Right). The distribution of RMS time differences from the 59 DOM pairs on string 21[62].

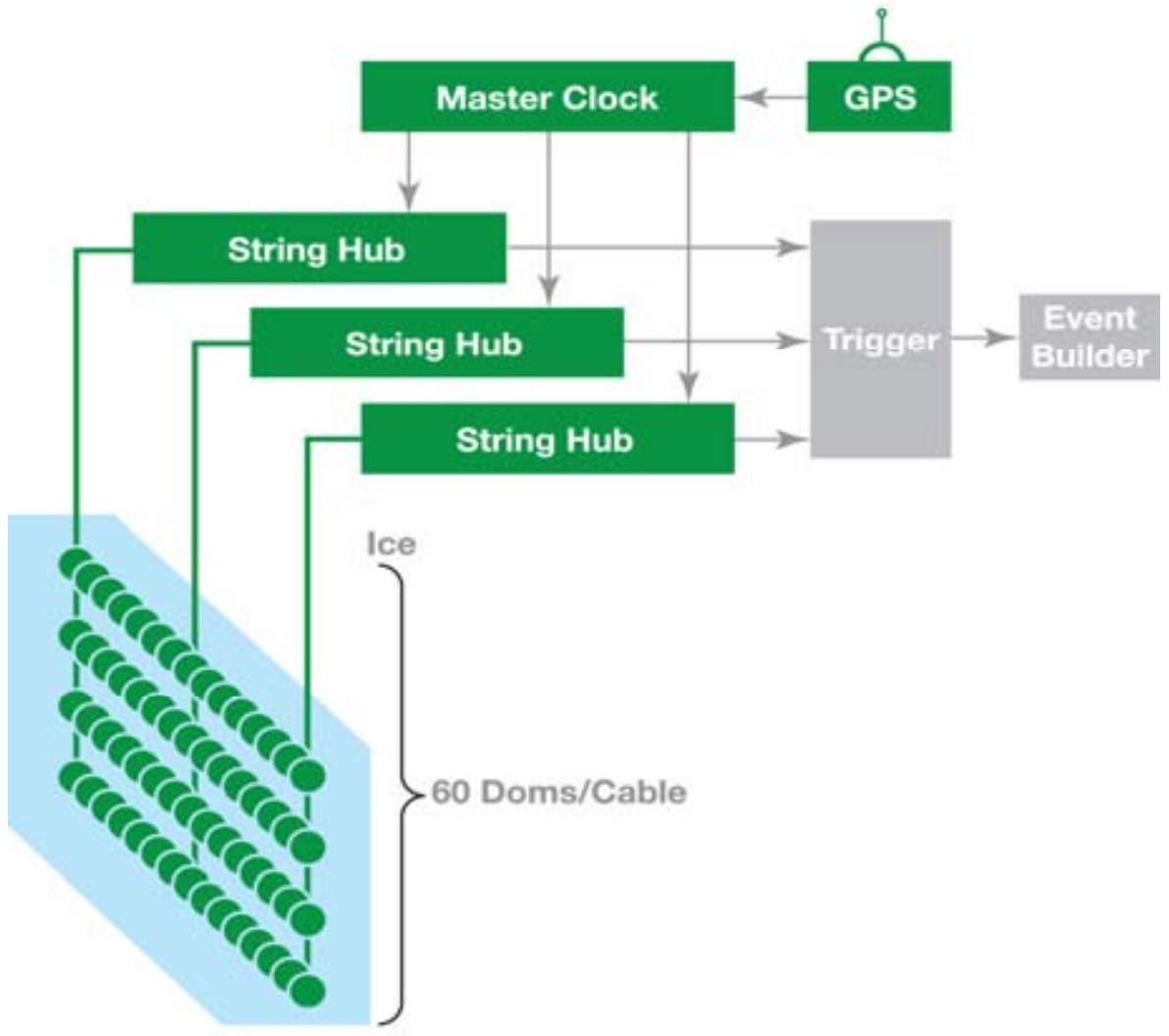

**Fig.19** A block diagram of the surface electronics. The string hub computers contain DOR cards which receive the signals from the DOMs. They pass these signals on to the trigger; hits within (typically) ±10 μs of a trigger are sent to the event builder, to be saved.

Besides sending control commands to the DOMs and receiving data, the DOR cards also distribute the ±48 V DC power to the DOMs. The cards also monitor the power consumption and communication error rates, and can turn individual twisted pairs on or off.

The String Hubs convert the time stamps from the DOMs into calibrated times, time-order these times, and send that information to a trigger processor. Trigger decisions are made on the basis of HLC hit times; SLC hits and the amplitude and waveform information are not used. A GPS receiver provides a single ‘master clock’ signal which is distributed throughout the surface electronics; matched cables are used to maintain timing across the system.

IceCube uses several trigger criteria[13]. The most commonly used trigger selects time intervals where eight DOMs (with local coincidences) fired within 5 μs. This collects most of the neutrino events. In 2008, a string trigger was added; it selects time intervals when five out of seven adjacent DOMs fired within 1.5 μs. This trigger has improved sensitivity for low energy events, especially upward-going muons. A proposed ‘topological’ trigger will be optimized for low-energy horizontal muons. Other triggers are under development for DeepCore.

When a trigger occurs, all data within a ±10 μs trigger window is saved, becoming an event. If multiple trigger windows overlap, then all of the data from the ORed time intervals are saved as a single event.

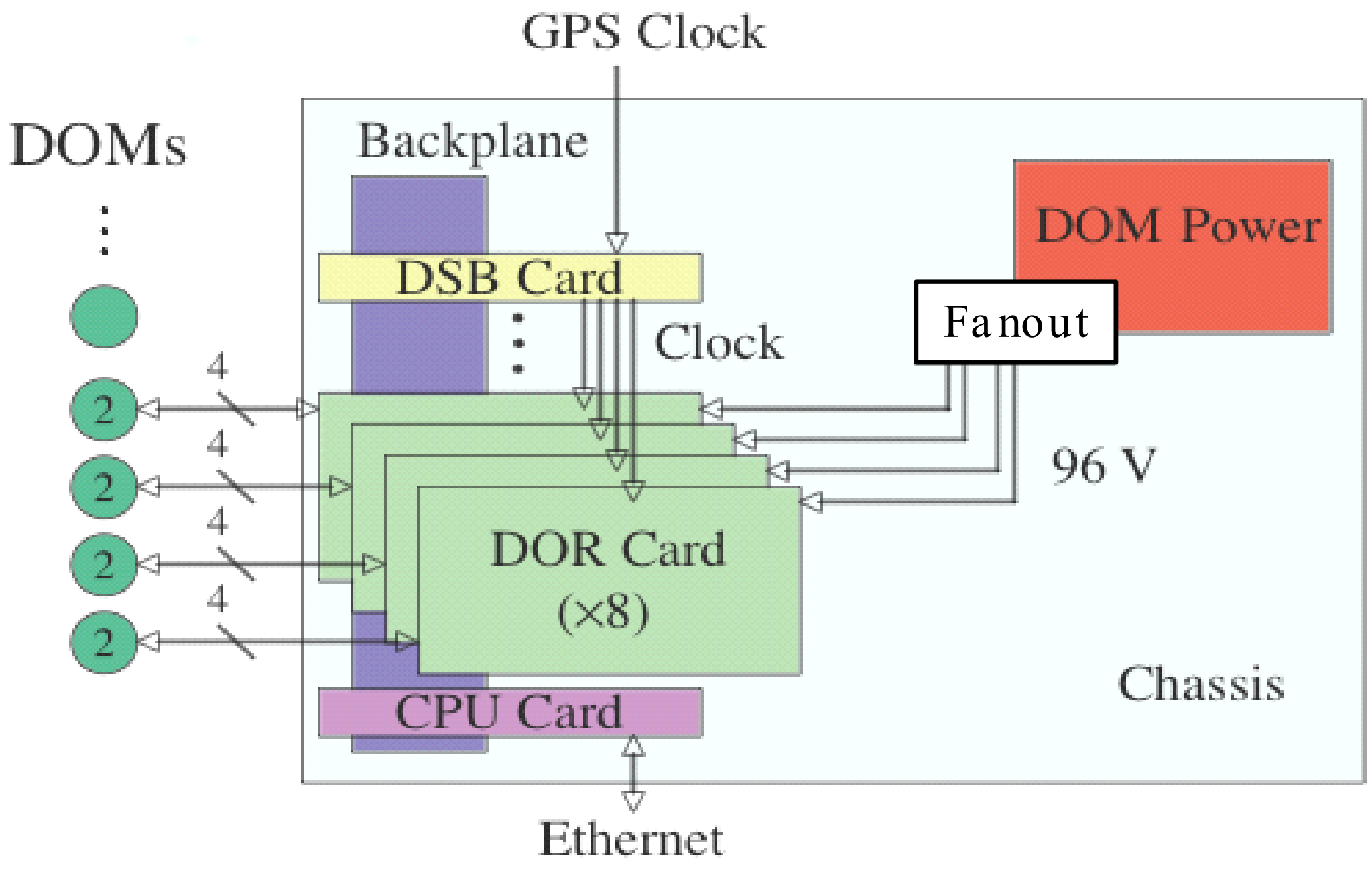

**Fig.20** A block diagram of a String Hub, showing the interfaces to the DOR cards.

IceTop uses two different trigger criteria, based on the number of hit stations. A station is a pair of nearby tanks. A station is considered hit if the high-gain DOM fired in one tank, in coincidence with the lower-gain DOM in the other. This was implemented in hardware by cross-wiring the local-coincidence circuitry. Higher energy events (above about 300 TeV) were collected with a trigger that required 8 hit stations; a prescaled lower energy trigger requires 3 stations to be hit.

All of the triggered data is reconstructed by an on-line filter system, and selected events are transmitted via satellite to the Northern hemisphere[64]. The filters use simple physics-based criteria, 'first-guess' reconstruction algorithms and simplified maximum likelihood fitting. Current filters select upward-going muons, cascades ($\nu_e$, $\nu_\tau$ and all-flavor neutral current interactions), extremely high-energy events, starting and stopping events, and air showers seen in IceTop. For the 40-string running, these filters selected about 6% of the events, comprising about 32 Gbytes/day. All of the data, including the data selected for satellite transmission, is stored on tapes at the South Pole station. The tapes are sent north during the Austral summer.

**Event Reconstruction**

The first stage of event reconstruction converts the PMT waveforms into photon arrival times, as is shown in Fig.21. The first step is to calibrate the waveform, converting ADC counts and ATWD fADC time bins into absolute times and voltages. The next step is to extract photon arrival times. This is done with several methods; the 'standard' approach is to perform a Bayesian peak unfolding; the algorithm searches for PMT-like pulses (with the correct shape), and removes them from the waveform, one by one.

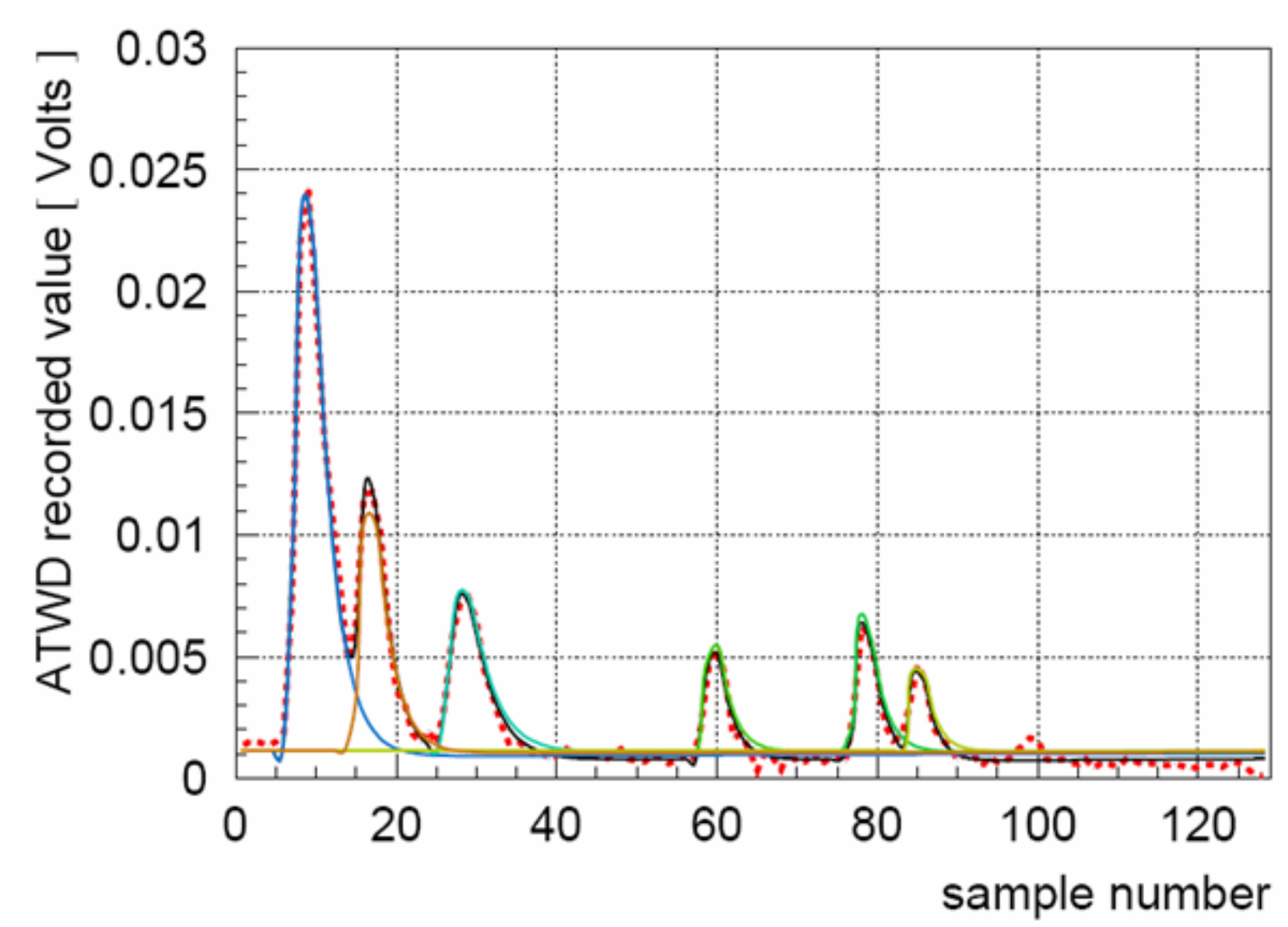

**Fig.21** The ATWD digitizer output from a typical event; multiple photoelectrons are clearly visible. Each time sample is 3.3 ns. The waveform is decomposed into a list of photon arrival times, which is used for event reconstruction[65].

These photon arrival times are used in maximum-likelihood fitting event reconstruction. IceCube can reconstruct the three different neutrino flavors based on the event topology. Fig.22 shows examples of three different types of interactions.

The top panel shows a kilometer-long muon track (or multiple parallel muons from a shower) traversing the detector. The long lever arm provides good directional reconstruction, better than 1 degree. The muon energy can be estimated by the track length (for muons which start and stop in the detector) or from the specific energy loss; at energies above 1 TeV, muon energy loss (*dE/dx*) is proportional to the muon energy.

Fig.22 (middle) shows a cascade from a simulated $\nu_e$ event. The light is nearly pointlike. Although most of the light is emitted near the Cherenkov angle, many of the photons scatter before being detected, partially washing out the angular information.

Fig.22 (bottom) shows a simulated few-PeV $\nu_\tau$ interaction forming a classic 'double-bang' topology. One 'bang' occurs when the $\nu_\tau$ interacts. That interaction also produces a $\tau$, which travels a few hundred meters before decaying, and producing a second 'bang.' Several other $\tau$ decay modes are under study in IceCube.

Other topologies are also of interest. A $\nu_\mu$ can interact in the detector, producing a hadronic shower from the struck nucleus, plus the $\mu$ track. If the neutrino interaction vertex can be clearly identified, it may be possible to search for neutrino sources above the horizon at moderate energies. Stopping muons may also be visible. Upward-going track pairs are also possible; these are predicted in some supersymmetric and Kaluza-Klein models[66].

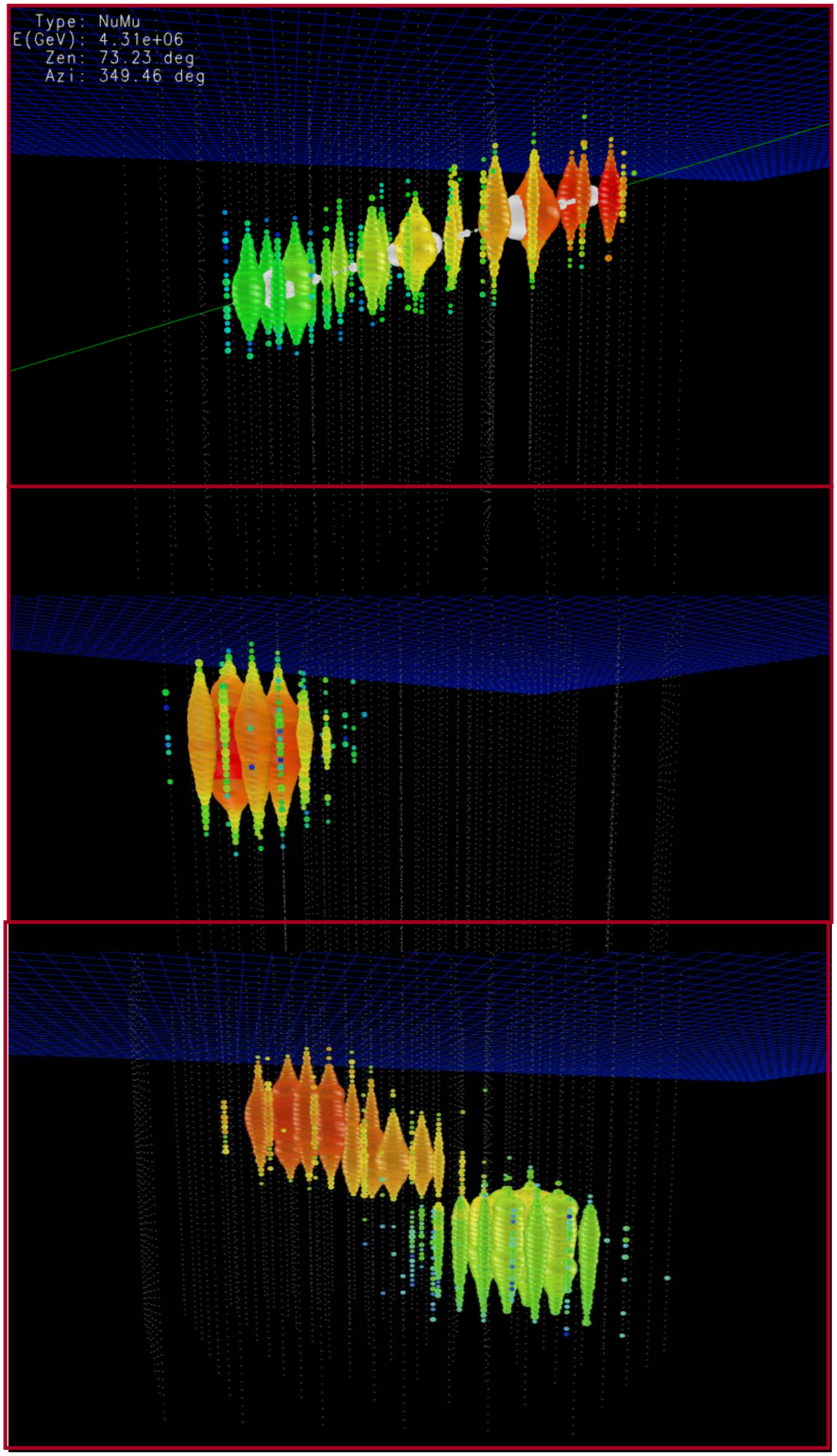

(a)

(b)

(c)

**Fig.22** Simulated events of the three types of neutrino interactions in IceCube: (a) $\nu_\mu$N --> $\mu$X (top), (b) $\nu_e$ N --> cascade (middle), and (c) a double bang, from $\nu_\tau$N --> $\tau$ cascade$_1$ --> cascade$_1$cascade$_2$ (bottom). Each circle represents one active optical module; the size of the circles shows the number of detected photons, while the color represents the time, from red (earliest) to blue (latest). In the top panel, the white shows the stochastic muon energy deposition along its track[13].

Of course, the vast majority of triggers are down-going muons from cosmic-ray air showers. They outnumber neutrino-induced events by about 500,000:1. Rejection of this background is a major challenge for event selection.

Events are reconstructed by fitting them to one of these topological hypotheses. The likelihood fits are seeded with a variety of 'first guess' methods to find starting points. For muons, the main first guess method fits a moving plane to the launch times in the DOMs[67]. For a muon, the plane should have a velocity near the speed of light. An alternate approach uses the measured charge deposition along the 'long axis' of events such as in Fig.22 (top).

The maximum likelihood fitter calculates the likelihood for different track positions and directions, and, optionally, energy. It does this by using functions which account for the light propagation through the ice. These functions are calculated via a Monte Carlo simulation, which tracks individual photons through the ice, accumulating the amplitude and time information in a 7-dimensional histogram[68]. These tables give the probability distribution for a photon radiated from a track with a given orientation to reach a DOM at a given perpendicular distance and orientation as a function of time. Depth is one dimension; the depth-dependent optical properties of the ice are properly included.

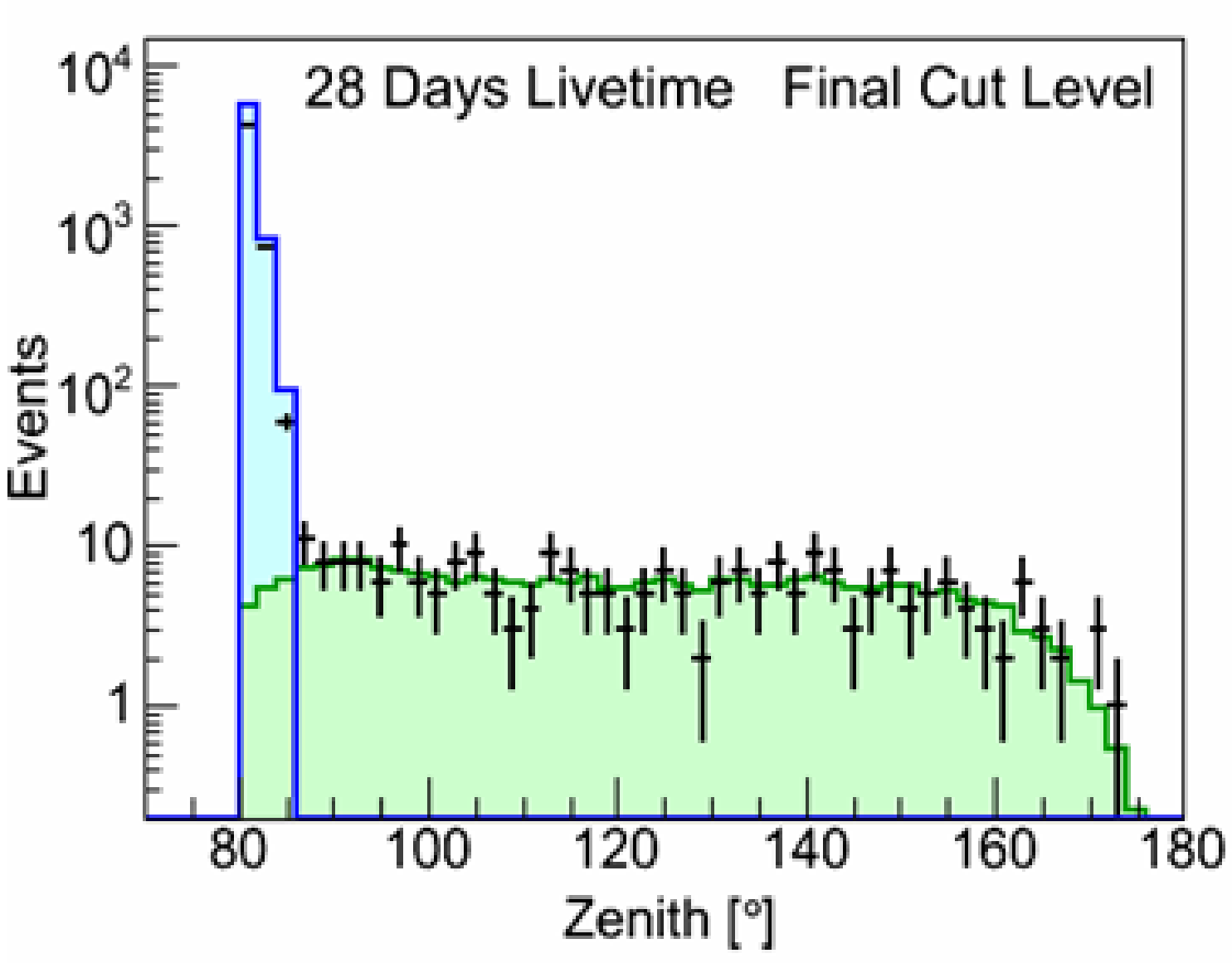

**Fig.23** The azimuthal angle for downward-going, or near downward-going muons in IceCube 22-string data, after tight cuts, compared with the results of cosmic-ray muon (blue) and neutrino (green) simulations. The coincident muon background is largely eliminated (4 downward-going events expected) and not shown here. From Ref. 13.

The huge background of downward-going muons must be eliminated using very tight cuts. In addition to the obvious cut on muon zenith angle, cuts are made based on the estimated errors returned from the likelihood fit. These probe the depth of the minimum in the likelihood function. Bayesian reconstructions are also used; the likelihood of a track having a given zenith angle is weighted by the relative size of the signal at that zenith angle. This effectively requires that the upward-going hypothesis be much more likely.

IceCube is large enough that there is a significant background from coincident muon events, when two (or more) muons from different cosmic-ray air showers are coincident in time in IceCube. Specific algorithms have been developed to find and reject these events.

These cuts leave a relatively clean sample of well-reconstructed neutrino events, as is shown in Fig.23. An irreducible background of atmospheric neutrinos remains. With the full detector, we expect to detect about 200 atmospheric $\nu_\mu$ interactions per day[69]. We will now discuss performance metrics and results from the partially completed detector; data with 22 strings was collected in 2007, and 40-string data was taken in 2008.

**Performance of IceCube**

Based on data taken with 40 strings, the expected effective area of the completed IceCube detector is shown in Fig.24[36]. The effective area is 2 to 3 times larger than had been anticipated in the original design[11, 12]. The main reasons for the improved efficiency are the unexpected optical quality of the ice in the lower half of the detector and the improved analysis methods exploiting the superior information provided by the DOMs.

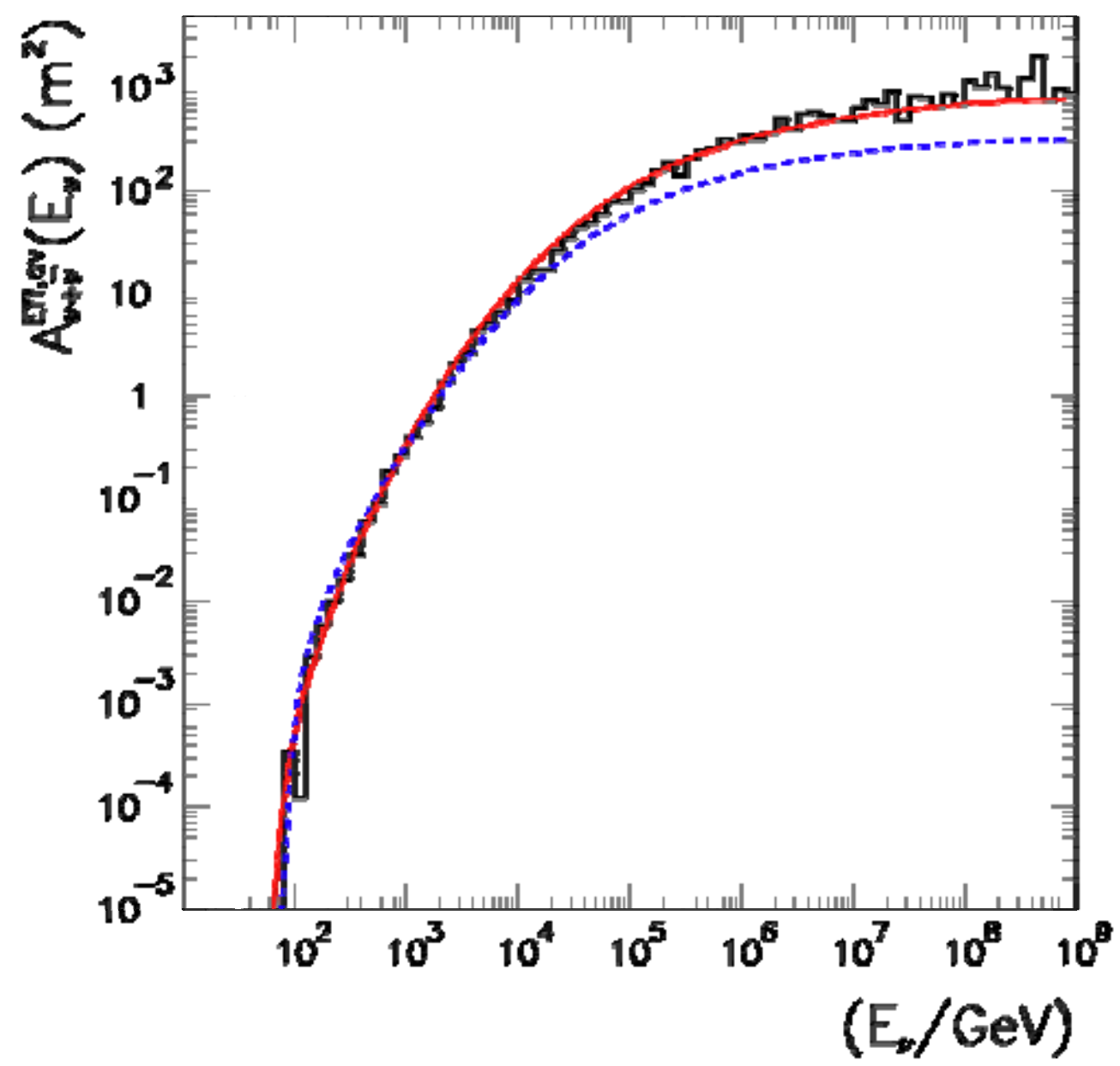

**Fig.24** The neutrino effective area (averaged over the Northern Hemisphere) from IceCube simulation (black histogram) is compared to the convolution of the approximate muon effective area[36] (solid red line) that is used in the estimates of event rates throughout this paper. The neutrino area is larger than the design area (shown as the dashed blue line) at high energy.

We have also performed a first test of the angular resolution for reconstructing muon tracks by observing the shadow of the Moon[70]. The Moon blocks cosmic rays from a 0.5-degree spot in the sky, reducing the flux of muons produced by cosmic rays. With 8 lunar months of data taken with 40 strings, we have observed a deficit of more than 5σ in the direction of the Moon;

see Fig.25. The next stage of this analysis will allow us to verify the IceCube angular resolution, which we anticipate to be close to 0.5 degree. The result confirms studies of the alignment of IceCube muon and IceTop shower directions. The accumulating data will allow us to actually check the angular resolution of the detector.

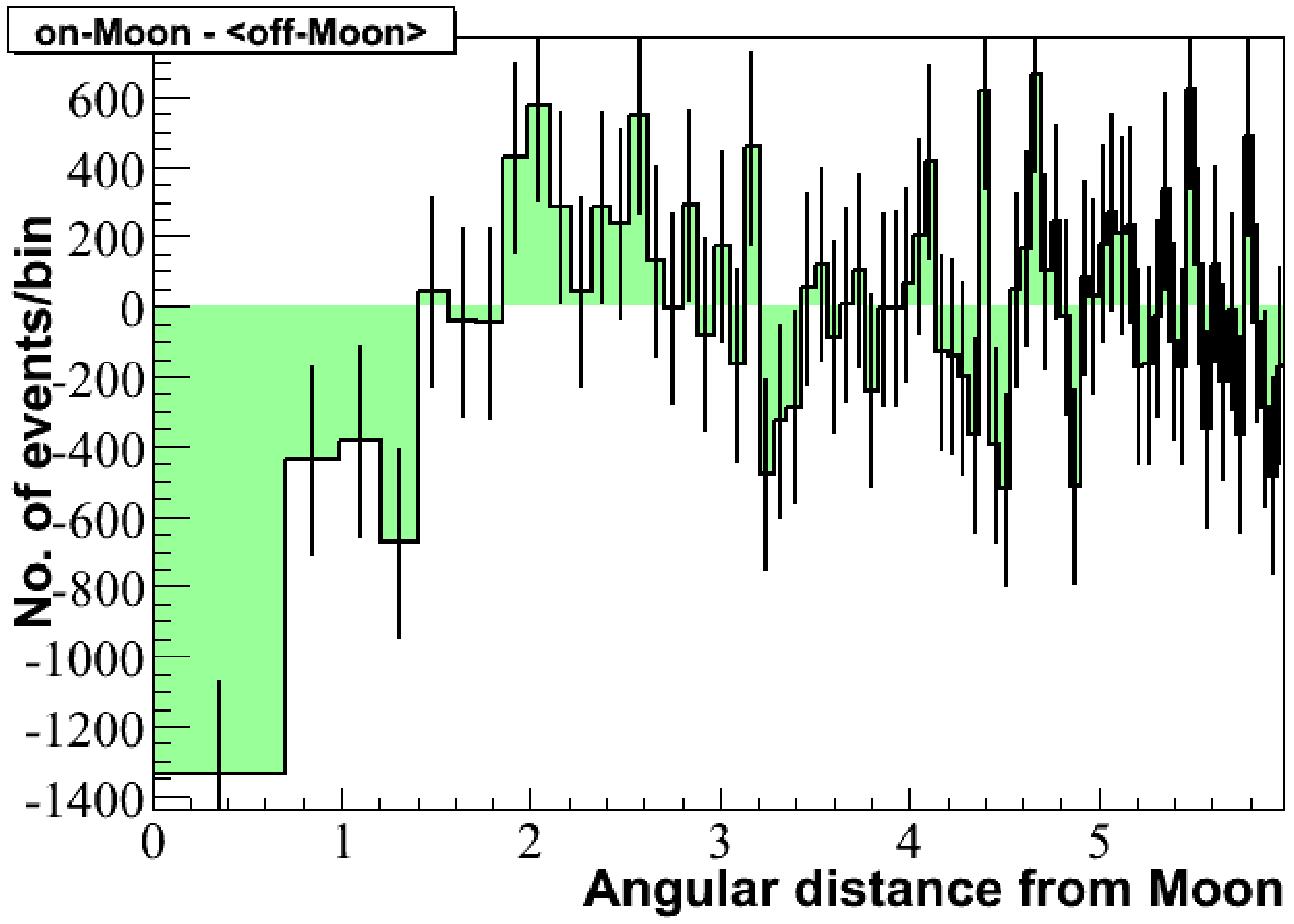

**Fig.25** Deficit of cosmic-ray muons in the direction of the Moon. Cosmic rays are blocked by the Moon, creating a shadow of one-half degree in the IceCube sky map. The shadow is visible as a deficit of secondary muons from cosmic-ray interactions in the atmosphere. The more-than-5σ deficit of events in the 40-string data confirms the pointing accuracy of the telescope[70].

The present status of the search for point sources of cosmic neutrinos is shown in Fig. 26. The neutrino map is the result of a novel unbinned search, a method that doubled the sensitivity of IceCube over the binned methods previously used[71]. In any given direction in the sky, two likelihoods are compared: i) that the data sample is consistent with a uniform background of atmospheric neutrinos and muons, modeled by the data; and ii) that the data reveal a point source in that particular direction. For modeling the possibility of a point source, we use a simulation that uses the actual point spread function of the detector and assumes an energy distribution, $E^{-2}$ for the map shown. In this way, potential sources of cosmic neutrinos are not only identified by the fact that they cluster in arrival direction; the analysis also takes into account that their reconstructed energy is large and less likely to be accommodated by the atmospheric background of relatively low-energy events. The latter is modeled by the data themselves.

Until recently, point-source searches with the IceCube neutrino telescope have been restricted to the Northern Hemisphere. One exclusively selects upward-going events as a way of rejecting the atmospheric muon background. Thus, one searches for cosmic sources in a relatively pure sample of atmospheric neutrinos. However, by preferentially selecting high-energy events, IceCube has sensitivity to high-energy sources over the full sky[72]. Above the horizon, the background consists of high-energy atmospheric muons or muon bundles, rather than neutrinos but the method still applies. Efficient energy estimators are now crucial for background rejection through rising energy thresholds above the horizon. Signal efficiency depends strongly on declination and effectively defines a lower energy threshold rising from the TeV energy in the North to PeV energies in the South for an $E^{-2}$ spectrum.

Unfortunately, with one half-year of data taken with a ½ $km^3$ detector, we are not yet sensitive to the predictions for cosmic neutrino fluxes associated with cosmic-ray sources previously discussed. However, with a growing detector, we expect to reach the required neutrino exposure within a few years.

If many sources contribute to the neutrino flux, then searches for a diffuse neutrino flux may be more sensitive than searches for individual point sources. The diffuse flux will be separable from the atmospheric background via several features. The first is its energy spectrum; a diffuse flux is expected to have a $dN/dE_\nu \sim E_\nu^{-2}$ energy spectrum, whereas the bulk of the atmospheric spectrum is much softer, $dN/dE_\nu \sim E_\nu^{-3.7}$. Prompt neutrinos, from the decay of charmed and heavy quarks, have a lower flux, but a harder spectrum, $dN/dE_\nu \sim E_\nu^{-2.8}$; at energies above about a few hundred TeV, prompt neutrinos will dominate the atmospheric background[73]. However, their spectrum is still softer than the diffuse flux. The second separator is the flavor spectrum; diffuse neutrinos will have travelled long distances, so the neutrino flux ratio should be $\nu_e:\nu_\mu:\nu_\tau = 1:1:1$. In contrast, the atmospheric flux is expected to be $\nu_e:\nu_\mu:\nu_\tau = 1:2:0$, while the prompt flux should be $\nu_e:\nu_\mu:\nu_\tau = 1:1:0$. This is why studies of multiple flavors are so important. IceCube is very close to being sensitive to the Waxman-Bahcall limit[40] with both $\nu_\mu$[74] and $\nu_e$[75].

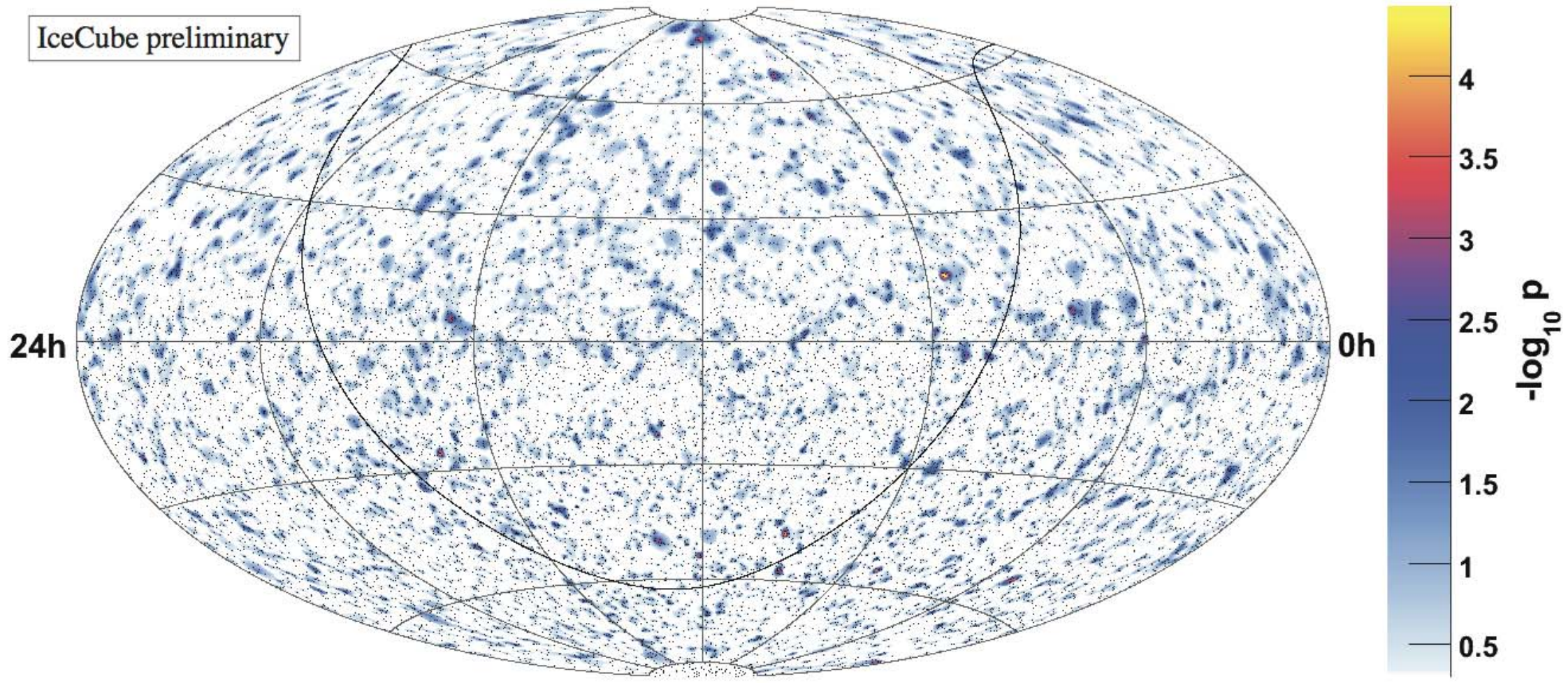

**Fig.26** Using declination and right ascension as coordinates, the map shows the probability for a point source of high-energy neutrinos with energies not readily accommodated by the steeply-falling atmospheric neutrino flux. Their energies range from 100 GeV to several 100 TeV. This map was obtained by operating IceCube with 40 strings for half a year[76]. The "hottest spot" in the map has an excess of 7 events, an excursion from the atmospheric background with a probability of $10^{-4.4}$. After taking into account trial factors, the probability to get a spot this hot somewhere in the sky is not significant. The map contains 6,796 neutrino candidates in the Northern Hemisphere and 10,981 down-going muons rejected to the $10^{-5}$ level in the Southern Hemisphere, shown as black dots.

## V. Other IceCube Science

Over a decade, IceCube will collect of order one million atmospheric neutrino events, covering the energy range 0.1 ~ $10^5$ TeV. This sample is two orders of magnitude larger than the total sample collected by AMANDA. Cosmic beams of even higher energy may exist. The sampling of physics topics ranges from the relatively straightforward to the positively exotic.

Even in the absence of new physics, a measurement of the neutrino cross-section at EeV energy represents a powerful test of the Standard Model. These interactions resolve partons with fractional momentum (Bjorken x-values) as low as $10^{-8}$. On the more exotic side, very-high-energy, short-wavelength neutrinos may interact with the space-time foam predicted by theories of quantum gravity. They will propagate through space like light through a crystal lattice and be delayed, with the delay depending on the energy [77]. This will appear to the observer as a violation of Lorenz invariance. Back-of-the-envelope calculations are sufficient to show that observations of neutrinos produced by gamma-ray bursts reach Planck-scale sensitivity.

In the end, the possibilities are only limited by our imagination and are still being identified. IceCube has contributed to glaciology[54] and monitors the South Pole atmosphere, including the ozone hole, using atmospheric muons[78]. One idea under study is to perform neutrino tomography of the Earth using atmospheric neutrinos[79]. The Earth is opaque to high-energy muon neutrinos because of their increased interaction cross sections; the diameter of the Earth represents one absorption length for a muon neutrino with an energy of 25 TeV. An initially uniform flux of atmospheric neutrinos from the Northern Hemisphere is modified in transit through the Earth; modifications are visible for neutrino energies in the 10-TeV region. Neutrinos within 30 degrees of vertical transit the Earth's core, whereas at larger angles, neutrinos traverse only the mantle.

### Beyond Astronomy

As the lightest of fermions and the most weakly interacting of particles, neutrinos occupy a fragile corner of the Standard Model, and one can realistically hope that they will reveal the first and most dramatic signatures of new physics; for a review, see Ref. 80. Some topics that IceCube will explore are listed below:

1. The search for signatures of the possible unification of particle interactions, including gravity, at the TeV scale. Neutrinos with energies approaching this scale would interact by gravity with large cross sections, similar to those of quarks and leptons. Their increased interaction cross-section will create dramatic signatures in a neutrino telescope including, possibly, the production of black holes.

2. The search for deviations from the neutrino's established oscillatory behavior that result from non-standard interactions, e.g. neutrino decay or quantum decoherence.

3. The search for a breakdown of the equivalence principle as a result of non-universal interactions with the gravitational field of neutrinos of different flavors.

4. Similarly, the search for breakdown of Lorentz invariance resulting from different limiting velocities of neutrinos of different mass. With energies of $10^3$ TeV and masses of order $10^{-2}$ eV or less, even the atmospheric neutrinos observed by IceCube reach Lorentz factors of $10^{17}$ or larger[77].

5. The search for particle emission from cosmic strings or any other topological defects or heavy cosmological remnants created in the early Universe. It has been suggested that they may be the sources of the highest-energy cosmic rays.

6. The search for magnetic monopoles, Q-balls and the like.

With its lower energy threshold and 10-Mton instrumented volume, DeepCore will explore additional physics topics, especially involving atmospheric-neutrino oscillations. A 10-GeV threshold will give us access to the first oscillation-induced $\nu_\mu$ flux minimum near 20 GeV, with unprecedented statistics. The low threshold opens the energy window for atmospheric-neutrino oscillation measurements, including $\nu_\mu$ disappearance, first observation of $\nu_\tau$ appearance and, possibly, if the mixing angle $\theta_{13}$ is large enough, the sign of the neutrino mass hierarchy[81].

**Galactic Supernova Explosions**

The IceCube/DeepCore detectors were designed to detect neutrinos with energies ranging from $10^{10}$ to $10^{21}$ eV. Nevertheless, a large burst of MeV supernova neutrinos streaming through the detector will produce an observable signal in the PMTs. Photons are predominantly produced by the Cherenkov radiation from showers produced by the interaction of supernova neutrinos $\bar{\nu}_e + p \to e^+ + n$ with protons in the ice. The Cherenkov radiation can be identified as a collective rise in the photomultiplier rates on top of their dark noise[82-83]. A Cherenkov photon is detected provided the neutrino interacts within 5.2 m of a digital optical module. This corresponds to a fiducial volume of ~ 2.5 megatons for 5,160 DOMs. The Cherenkov photons produce an excess counting rate above the steady 280 Hz dark noise of a photomultiplier deployed in the sterile Antarctic ice. The combined significance of the excess counts in 5,160 DOMs exceeds 5 $\sigma$ for a supernova collapse occurring as far as the Small Magellanic Cloud. IceCube will be able to provide a high-statistics measurement of the time profile corresponding to a 2.5-megaton conventional proton decay and supernova search experiment[84].

In a supernova search, IceCube simply counts neutrinos and does not observe the energy or direction of individual events. On the other hand, IceCube will collect over 1 million events from a supernova at 10 kpc, the most likely distance. For instance, a supernova explosion at the most probable distance of 10 kpc, releasing (after oscillations) 5x$10^{52}$ erg in electron antineutrinos with an average energy of 15 MeV, will produce a neutrino burst

$$N_\nu = \frac{1}{4\pi d^2}\frac{E_{tot}}{\langle E_\nu\rangle} = 1.75\times 10^{11}\,cm^{-2}\,\frac{E_{tot}}{5\times 10^{52}\,erg}\,\frac{15\,MeV}{\langle E_\nu\rangle}\left(\frac{10\,kpc}{d}\right)^2 \qquad (12)$$

This flux is 1.75 X $10^{11}$ per $cm^2$, and leads to the detection of 238 events per DOM for a total of $1.14x10^6$ events, i.e. over 1 million events sampled in time bins of 1.64 milliseconds:

$$N_{ev} = 1.14 \times 10^6 \left[ \frac{E_{tot}}{5 \times 10^{52}\, erg} \frac{15\, MeV}{\langle E_\nu \rangle} \left( \frac{10\, kpc}{d} \right)^2 \right] \left[ \frac{V_{\det}}{560\, m^3} \left( \frac{\langle E_\nu \rangle}{15\, MeV} \right)^3 \frac{N_{DOM}}{4800} \right] \quad (13)$$

The first square bracket scales the dependence on the number of neutrinos detected by IceCube according to the supernova parameters, the second according to the properties of the detector. The event rate is based on a flux of 183 produced Cherenkov photons for every MeV of neutrino energy. The rate has been determined from a variety of detailed simulations including GEANT.

As was the case for the historic 1987 observation, the high-statistics observations of a 21st Century supernova will further our understanding of star collapse and of neutrino physics, including the possible determination of $\theta_{13}$ and the mass hierarchy[71].

**The Search for Dark Matter**

The evidence that yet-to-be-detected, weakly interacting massive particles (WIMPs) make up dark matter is compelling. WIMPs are swept up by the Sun as the Solar System moves about the Galactic halo. Though interacting weakly, they will occasionally scatter elastically with nuclei in the Sun and lose enough momentum to become gravitationally bound. Over the lifetime of the Sun, enough WIMPs may accumulate in its center to establish equilibrium between their capture and annihilation. The annihilation products of these WIMPs represent an indirect signature of halo dark matter, their presence revealed by neutrinos that escape the Sun. The neutrinos are, for instance, the decay products of heavy quarks and weak bosons resulting from the annihilation of WIMPs into $\chi\bar{\chi} \to b\bar{b}$ or $W^+ + W^-$. These can be efficiently identified by IceCube because of the relatively large neutrino energy that is of order of the mass of the WIMP. The beauty of the indirect detection technique using neutrinos is that the astrophysics of the problem is understood. The source in the Sun has built up over solar time sampling the dark matter throughout the galaxy; therefore, any possible structure in the halo has been averaged out. Given a WIMP mass and interaction cross-section with ordinary matter, one can unambiguously predict the signal in a neutrino telescope. If not observed, the model is ruled out[85]. This is in contrast with indirect searches involving photons that are subject to theoretical uncertainties associated with the dark-matter distribution, especially in the very center of the Galaxy.

Although IceCube detects neutrinos of all flavors, sensitivity to neutrinos produced by WIMPs in the Sun is achieved by exploiting the degree accuracy with which $\nu_\mu$ can be pointed back to the Sun. The 22-string IceCube data have resulted in a limit on an excess flux from the Sun[86]. It improves on previous results by factors of 3 to 5 for WIMPs heavier than approximately 250 GeV. Though hardly competitive with direct searches for WIMPS with spin-independent interactions with ordinary matter, for spin-dependent interactions, IceCube has improved the best limits on WIMP cross sections by two orders of magnitude; see Fig.27.

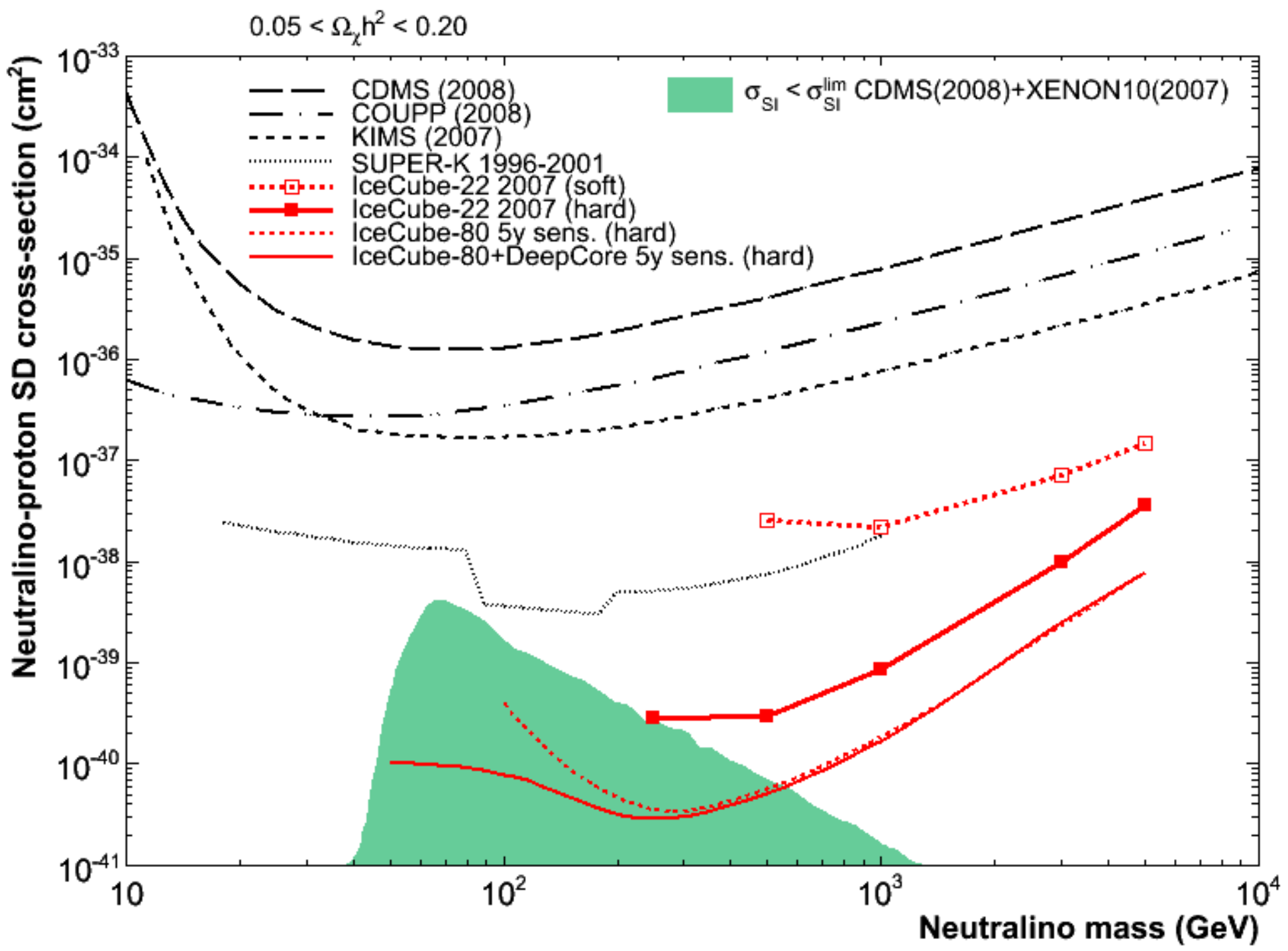

**Fig.27** The solid and dashed red lines show the 90%-confidence-level upper limits on the spin-dependent interactions of dark matter particles with ordinary matter[86]. The two lines represent extreme cases where neutrinos originate mostly from heavy quarks ("soft," top line) and weak bosons ("hard," bottom line) produced in the annihilation of dark-matter particles. Also shown is the reach of the complete IceCube and DeepCore with 5 years of data. The shaded area represents supersymmetric models not disfavored by direct searches for dark matter. Also shown are previous limits from direct experiments and from the SuperK experiment. The results improve by 2 orders of magnitude on the sensitivity previously obtained by direct experiments.

## Cosmic Ray Physics

The IceCube + IceTop combination is a potent cosmic-ray detector. IceTop detects showers, and IceCube observes the associated muons. The combination is sensitive to the cosmic-ray composition; heavier primary cosmic rays produce more muons at a given energy[87]. Also, uniquely, IceCube can search for muons hundreds of meters from the shower core; these muons come from high transverse-momentum interactions in the air shower[88].

By itself, IceCube will collect a huge sample of muons from cosmic-ray interactions; the homogeneity of the ice allows for careful studies of cosmic-ray arrival directions. Using 4.3 billion downward-going events, IceCube found a small anisotropy in the arrival directions of the cosmic rays, as is seen in Fig. 28[89]. The median muon energy is about 20 TeV; the primary energies are even higher. This is a puzzling result, as the arrival directions of charged particles should be scrambled by Galactic magnetic fields. This result complements earlier studies using Northern-Hemisphere detectors. Proposed interpretations fall into two categories: that the asymmetry in arrival directions of cosmic rays is either associated with unknown structure in the Galactic magnetic field, or with diffusive particle flows from nearby Galactic sources such as Vela.

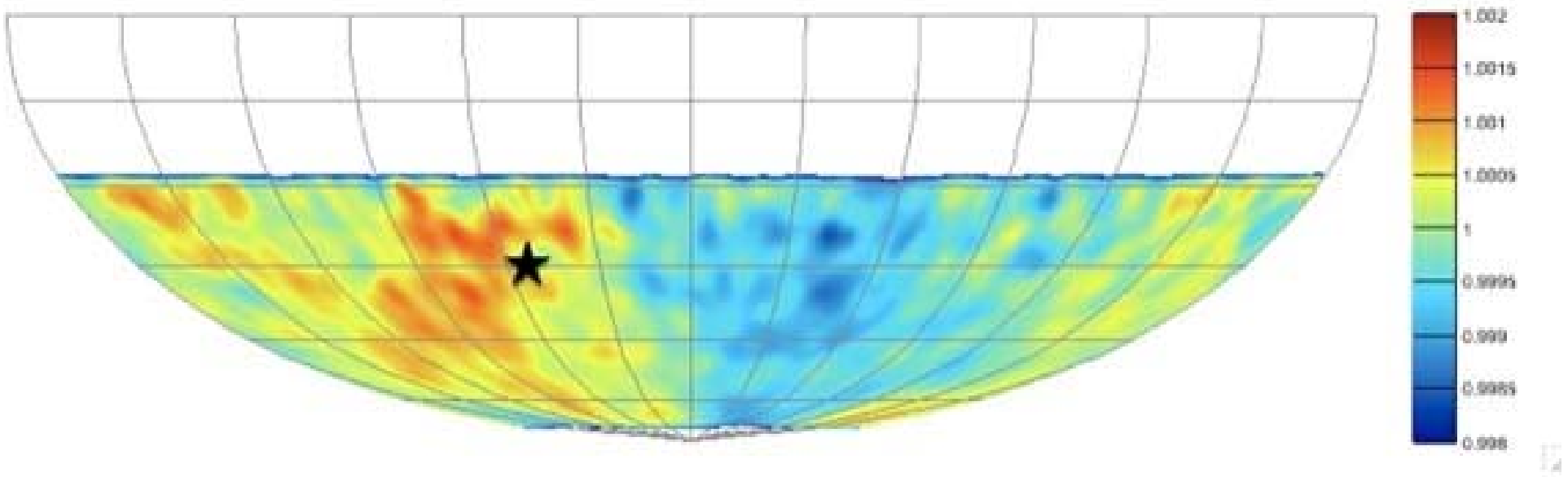

**Fig.28** The arrival direction of cosmic-ray muons detected with 22 IceCube strings. The color scale represents the relative intensity[79]. The star indicates the direction of Vela, the brightest gamma ray source in the sky.

The broad cosmic-ray anisotropy shown in the figure aligns with observations in the Northern Hemisphere[90]. It is intriguing that a prominent structure there seems to be associated with a major photon source, Geminga.

## Future Higher-Energy Developments

At energies above $10^{17}$ eV, a "guaranteed" source of neutrinos emerges. These are GZK neutrinos, produced when cosmic-ray protons with energies above $4\times10^{19}$ eV interact with the cosmic microwave background photons. The predicted rate for GZK neutrinos is at most of order 1 event/$km^3$ per year; IceCube is not big enough. One must trade threshold energy for active volume. Two techniques have been proposed for these large-area detectors: searching for radio waves or for acoustic pulses from neutrino interactions[2]. To be able to build a large enough detector with a reasonable number of elements requires that the signal waves (e.g. radio and acoustic pulses) have an attenuation length in the medium of order 1 km; this is a required (but not necessarily sufficient) condition to detect signals using an array with a detector spacing of order 1 km.

Radio pulses are generated from the charged particles that are produced in neutrino interactions[91]. High-energy electromagnetic showers contain about 20% more electrons than positrons because photons in the shower Compton scatter atomic electrons[92]. The Cherenkov radiation from the shower is coherent at wavelengths longer than the transverse size of the showers, i.e. above 20 MHz. The radio signal scales as the square of the neutrino energy, leading to an effective threshold of at least $10^{17}$ eV. Coherent Cherenkov radiation has been studied using beams of 25-GeV electrons striking ice, sand and salt targets. Measurements of the RF power, frequency spectrum and angular distributions are in good agreement with theoretical predictions[93].

In cold ice, radio-wave attenuation length is of order 1 km, far longer than for optical photons. The longer length allows 100 $km^3$ detectors to be built using a reasonable number of detection stations. A number of experiments are working to take advantage of this.

Most radio experiments have looked for signals from distant targets; the large separation between the radiator and the detector inevitably leads to higher thresholds, often above 1 EeV.

Most recently, the Antarctic Impulse Transient Antenna (ANITA) balloon experiment has twice circled Antarctica at altitudes around 35,000 m. Its 32 quad-ridged horn antennas scanned about $10^6$ $km^3$ of Antarctic ice[94].

Reaching a lower threshold requires placing the antennae in or very near the active volume. The first effort in this direction was by the Radio Ice Cherenkov Experiment (RICE) Collaboration, who installed dipole antennae in some of the AMANDA holes, and set limits down to $10^{17}$ eV[95]. A new proposal has been put forth to extend the IceCube array outward by placing antennae in shallow holes. This detector would ultimately cover 1,000 $km^3$ [96].

Unfortunately, recent measurements indicate that the acoustic attenuation length in polar ice is short, of order 200 m[97]; this will severely limit the effectiveness of an acoustic detector. Nevertheless, it may still be cost-effective to use acoustic detectors to supplement a radio array. Acoustic detectors have also been considered for other media, including salt domes, the Dead Sea and Siberian permafrost.

The Antarctic Ross Ice Shelf Antenna Neutrino Array (ARIANNA) collaboration has a new approach, using radio detectors on the 650-meter- thick Ross Ice Shelf in Antarctica[98]. The ice-water interface below the ice shelf is a near-perfect reflector for radio waves. With this reflection, radio waves from downward-going neutrinos will reach the surface, greatly increasing ARIANNA's sensitivity.

Although none of these experiments have observed a signal, their results are beginning to constrain models of GZK neutrinos. A neutrino experiment large enough to observe several GZK neutrinos per year would complement cosmic-ray experiments such as Auger. Unlike protons, these neutrinos point back to their sources. The reason is that a neutrino produced within a ~50-Mpc GZK radius from its source located at a typical cosmological distance of order Gigaparsecs, will reveal the source location within the relatively poor angular resolution of a neutrino telescope.

## VI. Conclusions

The 1 $km^3$ IceCube neutrino observatory detects Cherenkov radiation from charged particles produced in neutrino interactions. A total of 5,160 digital optical modules are being deployed on 86 vertical strings, with 60 DOMs attached at depths between 1,450 and 2,450 meters. The DOMs observe Cherenkov photons from charged particles produced in neutrino interactions. The bulk of IceCube is sensitive to neutrinos with energies above about 100 GeV; the DeepCore infill array may observe neutrinos with energies as low as 10 GeV. The IceTop surface array, located on the ice above IceCube, consists of 160 ice-filled tanks, each instrumented with two DOMs. It observes cosmic-ray air showers with a threshold of about 300 GeV.

IceCube acts like a tracking calorimeter, recording the pattern of energy deposition in the ice. Each DOM includes a complete data acquisition system. The higher pressures and low temperatures, along with the inaccessible locations, impose stringent requirements on these modules. Despite this, over 98% of the 4,740 deployed modules are working perfectly, with a global time resolution of about 2 ns.

Segmentation allows IceCube to separate $\nu_\mu$, $\nu_e$ and $\nu_\tau$ interactions. We have developed reconstruction methods that effectively separate upward-going muons from $\nu_\mu$ interactions from the much-more-intense cosmic-ray muon background. These methods achieve an angular resolution of better than 1 degree for long tracks.

The early data from IceCube is extremely promising, and the partial detector is now observing over 10,000 neutrino events per year.

We thank our IceCube collaborators for numerous discussions on neutrino physics. We thank Evelyn Malkus and Kim Krieger for help with graphics and proofreading. Juan Carlos Diaz-Velez for the event displays in Fig. 22. This work was funded in part by the National Science Foundation under grant number 0653266, OPP-0236449, PHY-0354776 and by the Department of Energy under contract number DE-AC-76SF-00098.